\documentclass[aps,pra,twocolumn,superscriptaddress,postprint]{revtex4-2}

\usepackage{amsfonts}
\usepackage{amssymb}
\usepackage{amsmath}
\usepackage[largesc, helvratio=0.96]{newtxtext}
\usepackage[varbb]{newtxmath}
\usepackage[scale=1.06]{newtxtt}
\usepackage[english]{babel}
\usepackage{tumcolors}
\usepackage{tummath}
\usepackage{siunitx}
\usepackage{tikz}
\usepackage[straightvoltages, europeanresistors]{circuitikz}
\usepackage{pgfplots}
\usepackage{pgfplotstable}
\usepackage{upgreek}
\usepackage{hyperref}
\usepackage[autostyle, english=american]{csquotes}
\usepackage{orcidlink}
\MakeOuterQuote{"}

\addto\captionsenglish{%
  \renewcommand{\figurename}{FIG.\@}%
}
\addto\captionsenglish{%
}

\newcommand{\figref}[2][{}]{Fig.~\hyperref[#2]{\ref{#2}#1}}
\newcommand{\tabref}[2][{}]{\tablename~\hyperref[#2]{\ref{#2}#1}}
\renewcommand{\imu}{\i}

\hypersetup{
  colorlinks = true,
  citecolor = TUMBlue,
  linkcolor = TUMBlue,
  urlcolor = TUMBlack,
}

\usetikzlibrary{arrows, decorations.markings, patterns, external,positioning}
\usepgfplotslibrary{units,fillbetween}
\usepgfplotslibrary{groupplots}%
\pgfplotsset{compat=newest}
\tikzexternaldisable%

\pgfplotsset{
  layers/axis lines on top/.define layer set={
      axis grid,
      pre main,
      axis background,
      main,
      axis ticks,
      axis tick labels,
      axis lines,
      axis descriptions,
      axis foreground,
    }{/pgfplots/layers/standard},
}

\DeclareSIUnit{\arbitraryunits}{arb.\,{}units}
\begin{document}

\title{Stochastic correction to the Maxwell-Bloch equations via the positive \texorpdfstring{$\bm{P}$}{$P$} representation}

\author{Johannes~\surname{Stowasser}\,\orcidlink{0009-0007-2240-163X}}
\affiliation{TUM School of Computation, Information and Technology, \href{https://ror.org/02kkvpp62}{\textcolor{TUMBlue}{Technical University of Munich}}, 85748 Garching, Germany\looseness=-1}
\author{Felix \surname{Hitzelhammer}\,\orcidlink{0009-0006-5335-7363}}
\affiliation{Institute of Physics, NAWI Graz, \href{https://ror.org/01faaaf77}{\textcolor{TUMBlue}{University of Graz}}, 8010~Graz, Austria}
\author{Michael~A.~\surname{Schreiber}\,\orcidlink{0000-0002-6003-3897}}
\affiliation{TUM School of Computation, Information and Technology, \href{https://ror.org/02kkvpp62}{\textcolor{TUMBlue}{Technical University of Munich}}, 85748 Garching, Germany\looseness=-1}
\author{Ulrich~\surname{Hohenester}\,\orcidlink{0000-0001-8929-2086}}
\affiliation{Institute of Physics, NAWI Graz, \href{https://ror.org/01faaaf77}{\textcolor{TUMBlue}{University of Graz}}, 8010~Graz, Austria}
\author{Gabriela~\surname{Slavcheva}\,\orcidlink{0000-0001-5474-9808}}
\affiliation{Institute of Physics, NAWI Graz, \href{https://ror.org/01faaaf77}{\textcolor{TUMBlue}{University of Graz}}, 8010~Graz, Austria}
\affiliation{Quantopticon, Chicago, Illinois 60615, USA}
\author{Michael~\surname{Haider}\,\orcidlink{0000-0002-5164-432X}}
\email[Contact author:~]{michael.haider@tum.de}
\homepage[]{\newline https://www.ee.cit.tum.de/cph}
\affiliation{TUM School of Computation, Information and Technology, \href{https://ror.org/02kkvpp62}{\textcolor{TUMBlue}{Technical University of Munich}}, 85748 Garching, Germany\looseness=-1}
\author{Christian~\surname{Jirauschek}\,\orcidlink{0000-0003-0785-5530}}
\affiliation{TUM School of Computation, Information and Technology, \href{https://ror.org/02kkvpp62}{\textcolor{TUMBlue}{Technical University of Munich}}, 85748 Garching, Germany\looseness=-1}
\affiliation{TUM Center for Quantum Engineering (ZQE), 85748 Garching,
  Germany}


\begin{abstract}
  Focusing on two-level atoms, we apply the positive $P$ representation to a full-wave mixed bosonic and fermionic system of Jaynes-Cummings type and identify an advantageous degree of freedom in the choice of the involved nonorthogonal fermionic basis states. On this basis, we propose a stochastic correction to the Maxwell-Bloch equations by relating them to a stochastic differential equation on a nonclassical phase space, which captures the full second quantization dynamics of the system. This approach explores the connection between semiclassical and field-quantized treatments of light-matter interaction and can potentially be used for the simulation of nonclassical light sources while retaining the main advantages of a semiclassical model.

  \bigskip

  \noindent\copyright{}~2024 American Physical Society, Phys.\ Rev.\ A \textbf{110}, 013704 (2024)\\
  DOI:~\href{https://doi.org/10.1103/PhysRevA.110.013704}{\textcolor{TUMBlue}{10.1103/PhysRevA.110.013704}}
\end{abstract}

\maketitle

\section{\label{sec-introduction}Introduction}

Nonclassical light is expected to enable breakthrough applications in emerging quantum technologies, such as quantum computing and quantum simulation, as well as quantum sensing and metrology~\cite{blais_cavity_2004,gross_quantum_2017,todorov_thz_2024,rundle_overview_2021}. Therein, effects like photon antibunching, entanglement, and squeezing play a significant role~\cite{pathak_classical_2018}. On the other hand, field-quantized numerical simulations of the light-matter interaction in, e.g., optoelectronic devices are challenging~\cite{georgescu_quantum_2014}. Realistic devices often exhibit considerable complexity, i.e., they may feature coherent interactions of many optical modes with a large number of atomic systems, as well as additional incoherent decay mechanisms and losses. Hence, straightforward approaches using orthogonal basis state expansion or moment recursion rapidly encounter computational hardware limitations or excessive simulation durations, primarily due to exponential scaling issues~\cite{barry_qubit_2008}.

Semiclassical simulations have a greatly reduced numerical load and provide a viable alternative to full quantum modeling of optoelectronic devices such as quantum cascade~\cite{freeman_laser-seeding_2013,jaidl_comb_2021,jirauschek_optoelectronic_2019,popp_modeling_2024,seitner_theoretical_2024,seitner_backscattering-induced_2024} and quantum dot (QD)~\cite{bardella_self-generation_2017,cartar_self-consistent_2017,majer_cascading_2010,slavcheva_model_2008,slavcheva_ultrafast_2019} lasers in the classical optical field limit. However, in the case of devices like single-photon sources based on QDs~\cite{hanschke_quantum_2018,trivedi_generation_2020}, which depend on the quantization of the optical field, these semiclassical methods fall short in delivering an accurate description. For comparisons between semiclassical and full quantum models, see, e.g., \cite{chen_understanding_2019,li_comparison_2019,waks_cavity_2010}. Semiclassical formalisms have been stochastically enhanced to include spontaneous emission~\cite{ikeda_theory_1980,jirauschek_dynamic_2023,roos_spontaneous_2023,slavcheva_fdtd_2004}, a nonclassical feature which is useful for the investigation of noise in lasers~\cite{dutra_maxwell-bloch_2000,popp_modeling_2024}, laser line widths~\cite{cerjan_quantitative_2015,pick_ab-initio_2015}, random lasers~\cite{andreasen_inherent_2012,andreasen_numerical_2010}, and active metamaterials~\cite{fang_modeling_2021,pusch_coherent_2012}. This has been achieved by adding specific noise terms to the widely used Maxwell-Bloch (MB) equations (see~\cite{allen1987optical,jirauschek_optoelectronic_2019}). Superfluorescence~\cite{andreasen_finite-difference_2009,benediktovitch_quantum_2019,chuchurka_stochastic_2023,chuchurka_stochastic2_2023,drummond_quantum_1991}, optical bistability~\cite{ben-aryeh_quantum_1989}, and thermal cavity noise~\cite{andreasen_finite-difference_2008} have also been statistically modeled in the MB framework by adding decay-induced fluctuations or in-coupled black-body radiation in the latter case. The fluctuation statistics up to second order can be derived by the Heisenberg-Langevin method, where expressions for the correlations are obtained by the generalized Einstein relations and the fluctuation-dissipation theorem~\cite{scully_quantum_1997,louisell_quantum_1973}. Intuitively, this type of noise is colored by the quantum system but does not account for the inherent quantum noise, which is independent of the decay in an open system (see~\cite{breuer_theory_2002}) and instead has its roots in the noncommutative nature of quantum mechanics itself~\cite{gardiner_quantum_2004}.

To address this limitation, we adopt the positive $P$ representation~\cite{drummond_generalised_1980}, a quantum optical phase space method~\cite{rundle_overview_2021,walls_quantum_1994} which, under specific conditions, allows us to use classical statistical physics according to the quantum-classical correspondence~\cite{carmichael_statistical_2_2007}. The underlying probability distribution of the abstract phase-space variables is then governed by a Focker-Planck equation (FPE) and can be directly sampled by means of a corresponding stochastic differential equation (SDE)~\cite{carmichael_statistical_1_1999}. Since the positive $P$ representation was initially developed for purely bosonic systems, the literature dealing with SDEs for fermionic or mixed bosonic and fermionic systems \cite{barry_qubit_2008,chuchurka_quantum_2023,mandt_stochastic_2015,ng_real-time_2011,ng_simulation_2013,olsen_phase-space_2005}, as required for the treatment of light-matter interaction, is sparse. Indeed, the resulting SDEs are prone to suffer from inherent instabilities, nonlinearities, and singularities. Alternatively, a few other stochastic approaches applicable to fermions exist. These include Gaussian phase space representations~\cite{corney_gaussian_2003,corney_gaussian_2004,corney_gaussian_2006,rousse_simulations_2024}, Grassmann phase space methods~\cite{dalton_grassmann_2013,dalton_grassmann_2016,dalton_grassmann_2017,polyakov_grassmann_2016}, and operator algebra techniques involving characteristic functions (Haken-Risken-Weidlich method)~\cite{carmichael_nonlinear_1986,drummond_quantum_1991,gardiner_quantum_2004,haken_quantum_1967,jen_positive_2012}. Generally the last method, originating in the investigation of lasers, is correct in the small noise limit, which may not be applicable in the strong coupling regime of cavity quantum electrodynamics (cavity QED)~\cite{carmichael_statistical_2_2007}.

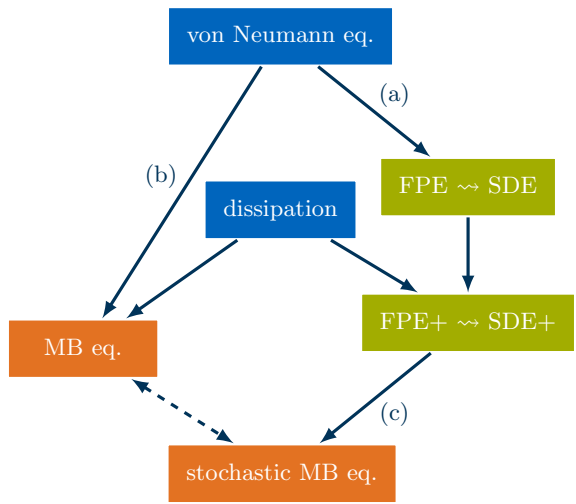
\begin{figure}[t]
  \centering
  \tikzexternalenable
  \begin{tikzpicture}
    \node[fill=TUMBlue, text=TUMWhite, draw=TUMWhite, thick, inner sep=2.5mm] (neumann) {\vphantom{Mg}von Neumann eq.};
    \node[below left= 3.35cm and 0.1cm of neumann, fill=TUMOrange, text=TUMWhite,
        draw=TUMWhite, thick, inner sep=2.5mm, minimum width=2cm] (mb) {\vphantom{Mg}MB eq.};
    \node[below =1.5cm of neumann, fill=TUMBlue, text=TUMWhite, draw=TUMWhite, thick, inner sep=2.5mm, minimum width=2cm] (dissipation) {\vphantom{Mg}dissipation};
    \node[below right=1.2cm and -0.2cm of neumann, fill=TUMGreen, text=TUMWhite, draw=TUMWhite, thick, inner sep=2.5mm, minimum width=2cm] (sde1) {\vphantom{Mg}FPE $\leadsto$ SDE};
    \node[below =1cm of sde1, fill=TUMGreen, text=TUMWhite,
        draw=TUMWhite, thick, inner sep=2.5mm, minimum width=2cm] (sde2) {\vphantom{Mg}FPE+ $\leadsto$ SDE+};
    \node[below left=1.2cm and -0.5cm of sde2, fill=TUMOrange, text=TUMWhite,
        draw=TUMWhite, thick, inner sep=2.5mm, minimum width=2cm] (stochasticmb) {\vphantom{Mg}stochastic MB eq.};

    \draw[TUMBlueDarker, very thick, -latex] (neumann) -- (sde1) node [midway,above right=-0.1cm] {(a)};
    \draw[TUMBlueDarker, very thick, -latex] (neumann) -- (mb) node [midway,above left=-0.1cm] {(b)};
    \draw[TUMBlueDarker, very thick, -latex] (sde1) -- (sde2);
    \draw[TUMBlueDarker, very thick, -latex] (dissipation) -- (sde2);
    \draw[TUMBlueDarker, very thick, -latex] (dissipation) -- (mb);
    \draw[TUMBlueDarker, very thick, -latex] (sde2) -- (stochasticmb) node [midway,below right=-0.1cm] {(c)};
    \draw[TUMBlueDarker, very thick, dashed, latex-latex] (mb) -- (stochasticmb);
  \end{tikzpicture}
  \tikzexternaldisable
  \caption{Outline of the derivation of the stochastic MB equations and their relation to the MB equations. Some of the key concepts involved are highlighted: (a) nonorthogonal basis state expansion (positive $P$ representation), (b) dipole approximation for the interaction Hamiltonian, (c) change of fermionic and bosonic variables.}
  \label{fig-overview}
\end{figure}
In this contribution, we show that some of the problems associated with mixed bosonic and fermionic systems can be remedied by engineering a set of diffusion-optimized nonorthogonal basis states for the expansion of the fermionic system component in terms of the positive $P$ representation. Moreover, by a multistep process, we derive stochastic partial differential equations (SPDEs)~\cite{gardiner_stochastic_2009,werner_robust_1997} for a dissipative two-level atom (TLA) in a perfect optical cavity featuring phase space variables with a direct physical interpretation, namely, inversion, coherence, as well as electric and magnetic fields. This corresponds to a stochastic correction to the MB equations, where the evolution of the deterministic part of the SPDE represents the standard MB equations. The outline of the aforementioned derivation process is shown in \figref{fig-overview}.

Previous approaches for the derivation of modified or corrected MB equations rely on, e.g., the Wigner quasi-probability distribution~\cite{hofmann_quantum_1999}, the orthogonal projection of single-atom cavity QED~\cite{mabuchi_derivation_2008}, a FPE in the bad cavity limit~\cite{wang_cavity_modified_1997}, or the It\^{o} formula applied to a suitable stochastic ansatz~\cite{chuchurka_stochastic_2023,chuchurka_stochastic2_2023}. Our stochastic correction to the MB equations by means of the inherent quantum noise supplied by the positive $P$ representation SDE goes beyond adding decay-induced fluctuations with second-order accurate statistics. In this way, we propose to integrate key quantum effects while maintaining compatibility with existing semiclassical solvers~\cite{oskooi_meep_2010,quantopticon_quantillion_2024,riesch_mbsolve_git_2024}, leveraging their numerical efficiency. This connection between stochastic modeling and proven simulation methods allows for a more intuitive understanding of light-matter interaction.

The paper is structured as follows: In Sec.~\ref{sec-models} we compare suitable existing second-quantization and semiclassical models for light-matter interaction in optical cavities. We apply the positive $P$ representation to the field-quantized model in Sec.~\ref{sec-sde} and state the associated full-wave Jaynes-Cummings-type SDE, introducing diffusion-optimized nonorthogonal fermionic basis states. The simulation of these SDEs poses some general and practical problems, which are discussed in Sec.~\ref{sec-simulation}. Ultimately, we want to simulate realistic optoelectronic devices. Therefore, we incorporate unavoidable dissipation into our SDE in Sec.~\ref{sec-dissipation}. Our main result, the stochastic correction to the MB equations, is given Sec.~\ref{sec-variables}.

\section{\label{sec-models}Light-matter interaction}

\begin{figure}
  \centering
  \tikzexternalenable
  \begin{tikzpicture}
    \pgfmathsetmacro{\l}{2.8};
    \pgfmathsetmacro{\h}{1.5};

    \pgfmathsetmacro{\cx}{\l+0.4};
    \pgfmathsetmacro{\cy}{\h-0.6};
    \pgfmathsetmacro{\cl}{0.5};
    \pgfmathsetmacro{\cdr}{0.1};

    \pgfmathsetmacro{\d}{0.2};

    \draw[draw=TUMBlue,thick] plot[domain=0:1,smooth] ({-\l+\x*2*\l},{\h*sin(pi*(\x r))});
    \draw[draw=TUMBlue,thick,dashed] plot[domain=0:1,smooth] ({-\l+\x*2*\l},{-\h*sin(pi*(\x r))});
    \draw[draw=TUMBlue,thick] plot[domain=0:1,smooth] ({-\l+\x*2*\l},{\h*sin(2*pi*(\x r))});
    \draw[draw=TUMBlue,thick,dashed] plot[domain=0:1,smooth] ({-\l+\x*2*\l},{-\h*sin(2*pi*(\x r))});
    \draw[draw=TUMBlue,thick] plot[domain=0:1,smooth] ({-\l+\x*2*\l},{\h*sin(3*pi*(\x r))});
    \draw[draw=TUMBlue,thick,dashed] plot[domain=0:1,smooth] ({-\l+\x*2*\l},{-\h*sin(3*pi*(\x r))});

    \node[text=TUMBlue,above=0.0cm] (omega1) at (0,\h) {$\omega_{1}$};
    \node[text=TUMBlue,above=0.0cm] (omega2) at (-\l/2,\h) {$\omega_{2}$};
    \node[text=TUMBlue,above=0.0cm] (omega3) at (-2*\l/3,\h) {$\omega_{3}$};

    \draw[thick,dashed] (-\l,0) to (\l,0);
    \begin{scope}
      \clip (-\l-\d,-\h) rectangle (-\l,\h);
      \draw[thick] (-\l,-\h) to (-\l,\h);
      \draw[thick,pattern={north east lines},pattern color=black] (-\l-2*\d,-\h-2*\d) rectangle (-\l+2*\d,\h+2*\d);
    \end{scope}
    \begin{scope}
      \clip (\l,-\h) rectangle (\l+\d,\h);
      \draw[thick] (\l,-\h) to (\l,\h);
      \draw[thick,pattern={north east lines},pattern color=black] (\l-2*\d,-\h-2*\d) rectangle (\l+2*\d,\h+2*\d);
    \end{scope}

    \node[below] (PECl) at (-\l-\d,-\h) {\small PEC};
    \node[below] (PECr) at (\l+\d,-\h) {\small PEC};

    \draw[draw=TUMOrange,fill=TUMOrange] (-2*\l/3,0) circle (0.1cm);
    \node[text=TUMOrange,above=0.12cm] (tla) at (-2*\l/3,0) {\small TLA};
    \node[text=TUMOrange,below=0.12cm] (x0) at (-2*\l/3,0) {$x_0$};

    \draw[-latex,thick] (\cx+\d,\cy) -- (\cx+\cl+\d,\cy) node[right=-0.1cm]{\small $x$};
    \draw[-latex,thick] (\cx+\d,\cy) -- (\cx+\d,\cy+\cl) node[above=-0.05cm]{\small $z$};
    \draw[draw=black,fill=white,thick] (\cx+\d,\cy) circle (0.1cm);
    \draw[rotate around={45:(\cx+\d,\cy)}] (\cx+\d,\cy-\cdr) -- (\cx+\d,\cy+\cdr);
    \draw[rotate around={45:(\cx+\d,\cy)}] (\cx-\cdr+\d,\cy) -- (\cx+\cdr+\d,\cy);
    \node[below left=-0.02cm] at (\cx+\d,\cy) {\small $y$};
  \end{tikzpicture}
  \tikzexternaldisable
  \caption{Two-level atom located at position $x_{0}$ in an otherwise empty optical cavity without mirror losses, i.e., we assume perfect electric conductors (PECs) on either end. The first three cavity modes with angular frequencies $\omega_{1}$, $\omega_{2}$ and $\omega_{3}$ are shown.}
  \label{fig-atom-cavity}
\end{figure}
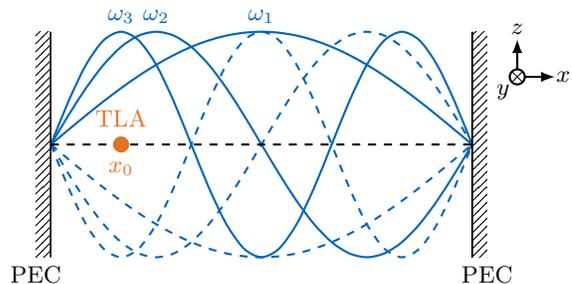

We consider one of the simplest possible systems for studying light-matter interaction: a TLA with lower level $\ket{\downarrow}$ and upper level $\ket{\uparrow}$ placed in a perfect optical cavity (see~\figref{fig-atom-cavity}). The energies of the levels are $-\hbar\Omega/2$ and $\hbar\Omega/2$ where $\Omega$ is the angular frequency of this transition and $\hbar$ denotes the reduced Planck constant.
We now look at two possible models, differing by whether the optical field is treated quantum mechanically or classically. The latter is only an approximation that precludes the treatment of all effects due to atom-field entanglement. This section sets the stage for our approach to the MB equations (see right branch of \figref{fig-overview}) and establishes a uniform notation that is used throughout the paper.

\subsection{\label{sec-quantization}The second quantization case}

The details of the optical field quantization in the perfect cavity depend on its geometrical features length $l$, transverse area $A$, and volume $V=lA$~\cite{meystre_elements_2007}. Let $x_{0}\in(0,l)$ be the position of the TLA in the cavity. We also need the vacuum speed of light $c=1/\sqrt{\mu_{0}\varepsilon_{0}}$ and the impedance of free space $Z=\sqrt{\mu_{0}/\varepsilon_{0}}$. Here $\varepsilon_{0}$ and $\mu_{0}$ denote the vacuum permittivity and permeability, respectively.

The $n$th cavity mode for $n\in\Nnums$ has the angular frequency
\begin{equation}
  \omega_{n}=\frac{\pi cn}{l}
\end{equation}
and wave number $k_{n}=\omega_{n}/c$. The electric field per photon in the cavity is given by
\begin{equation}
  \function{e_\mathrm{p}}{\omega_{n}}=\sqrt{\frac{\hbar\omega_{n}}{\varepsilon_{0}V}}\, ,
\end{equation}
and of course depends on the energy of the photon (by way of the angular frequency). Then the electric and magnetic field operators
\begin{equation}
  \begin{alignedat}{2}
     & \function{\op{E}_{z}}{x,t} &  & =\sum_{n=1}^{\infty}\function{e_{\mathrm{p}}}{\omega_{n}}\left[\function{\op{a}_{n}^{\dagger}}{t}+\function{\op{a}_{n}}{t}\right]\sin(k_{n}x)\, ,           \\
     & \function{\op{H}_{y}}{x,t} &  & =-\frac{1}{Z}\sum_{n=1}^{\infty}\function{e_{\mathrm{p}}}{\omega_{n}}\i\left[\function{\op{a}_{n}^{\dagger}}{t}-\function{\op{a}_{n}}{t}\right]\cos(k_{n}x)
    \label{eqn-field-operators-qm}
  \end{alignedat}
\end{equation}
can be written in terms of the bosonic creation $\op{a}_{n}^{\dagger}$ and annihilation $\op{a}_{n}$ operators of each cavity mode and the corresponding spatial mode functions $\sin(k_{n}x)$ and $\cos(k_{n}x)$. For later use, we introduce the notation
\begin{equation}
  \begin{alignedat}{2}
     & e_{n} &  & =\braket{\op{a}_{n}^{\dagger}+\op{a}_{n}}\, ,   \\
     & h_{n} &  & =\i\braket{\op{a}_{n}^{\dagger}-\op{a}_{n}}\, ,
    \label{eqn-en-hn}
  \end{alignedat}
\end{equation}
since we are mostly interested in the electric and magnetic fields
\begin{equation}
  \begin{alignedat}{2}
     & \function{E_{z}}{x,t} &  & =\sum_{n=1}^{\infty}\function{e_{\mathrm{p}}}{\omega_{n}}e_{n}(t)\sin(k_{n}x)\, ,             \\
     & \function{H_{y}}{x,t} &  & =-\frac{1}{Z}\sum_{n=1}^{\infty}\function{e_{\mathrm{p}}}{\omega_{n}}h_{n}(t)\cos(k_{n}x)\, ,
    \label{eqn-fields-qm}
  \end{alignedat}
\end{equation}
which are obtained by taking the expectation values of the respective operators.

Now directing our attention to the TLA, we introduce the pseudospin operators
\begin{equation}
  \begin{alignedat}{2}
     & \op{S}_{z} &  & =-\frac{1}{2}\ketbra{\downarrow}{\downarrow}+\frac{1}{2}\ketbra{\uparrow}{\uparrow}\, , \\
     & \op{S}_{+} &  & =\ketbra{\uparrow}{\downarrow}\, ,                                                      \\
     & \op{S}_{-} &  & =\ketbra{\downarrow}{\uparrow}={\op{S}_{+}}^{\dagger}\, .
  \end{alignedat}
\end{equation}
It is also common to use the operators $\op{\sigma}_{z}=2\op{S}_{z}$, $\op{\sigma}_{+}=\op{S}_{+}$, and $\op{\sigma}_{-}=\op{S}_{-}$~\cite{carmichael_statistical_2_2007} but we follow the same convention as~\cite{mandt_stochastic_2015}.

The second quantization description of the system leads to the Hamiltonian
\begin{equation}
  \begin{split}
    \op{H}_{\mathrm{JC}}= & \op{H}_{\mathrm{A}}+\op{H}_{\mathrm{F}}+\op{H}_{\mathrm{I}}                                                                                         \\
    =                     & \hbar\Omega\op{S}_{z}+\sum_{n=1}^{\infty}\hbar\omega_{n}\op{a}_{n}^{\dagger}\op{a}_{n}                                                              \\
                          & +\sum_{n=1}^{\infty}\hbar \function{g}{\omega_{n}}\sin(k_{n}x_{0})\left(\op{a}_{n}^{\dagger}+\op{a}_{n}\right)\bigl(\op{S}_{+}+\op{S}_{-}\bigr)\, .
  \end{split}
  \label{eqn-hamiltonian-jc}
\end{equation}
Here $\op{H}_{\mathrm{A}}$ and $\op{H}_{\mathrm{F}}$ are the free Hamiltonians of the atom and optical field, respectively. The interaction Hamiltonian $\op{H}_{\mathrm{I}}$ contains the frequency-dependent bosonic-fermionic coupling constants $\function{g}{\omega_{n}}$ and position factors $\sin(k_{n}x_{0})$. Note that \eqref{eqn-hamiltonian-jc} reduces to the analytically solvable Jaynes-Cummings model~\cite{larson_jaynes-cummings_2021} if we consider only one cavity mode and moreover drop the two energy non-preserving counterrotating terms $\op{a}\op{S}_{-}$ and $\op{a}^{\dagger}\op{S}_{+}$ in the interaction Hamiltonian, which corresponds to the rotating-wave approximation (RWA)~\cite{scully_quantum_1997}. An investigation of the RWA's influence on the TLA's stochastic dynamics can be found in~\cite{band_open_2015}.

\subsection{\label{sec-semiclassical}The semiclassical case}

If we treat the optical field classically, we can eliminate the bosonic degrees of freedom in the Hamiltonian \eqref{eqn-hamiltonian-jc} by introducing the dipole moment operator
\vspace{-1.5mm}
\begin{equation}
  \op{m}=m_{21}\op{S}_{+}+m_{21}^{\ast}\op{S}_{-}\, ,\label{eqn-dipole-operator}
  \vspace{-2mm}
\end{equation}
and using the dipole approximation
\vspace{-1.5mm}
\begin{equation}
  \op{H}_{\mathrm{I},\mathrm{DA}}=-\op{m}E_{z}
  \vspace{-2mm}
\end{equation}
for the interaction Hamiltonian. In this section, $\op{\rho}=\op{\rho}_{\mathrm{A}}$ denotes the atomic density operator, which fully determines the mixed state of the TLA, i.e., the only remaining quantum mechanical system. The corresponding matrix entries are
\vspace{-1.5mm}
\begin{equation}
  \begin{bmatrix}
    \rho_{11} & \rho_{12} \\
    \rho_{21} & \rho_{22}
  \end{bmatrix}=\begin{bmatrix}
    \braket{\downarrow|\op{\rho}|\downarrow} & \braket{\downarrow|\op{\rho}|\uparrow} \\
    \braket{\uparrow|\op{\rho}|\downarrow}   & \braket{\uparrow|\op{\rho}|\uparrow}
  \end{bmatrix}\, .\label{eqn-matrix-rho}
  \vspace{-2mm}
\end{equation}
Applications involving the simulation of realistic optoelectronic devices make it necessary to also account for scattering and dephasing~\cite{jirauschek_optoelectronic_2019}. For this reason, we consider the following dissipation superoperator in Lindblad form~\cite{lindblad_generators_1976,riesch_mbsolve_2021}
\vspace{-1.5mm}
\begin{equation}
  \begin{split}
    \mathcal{D}(\hat{\rho})= & \,r_{\mathrm{p}}\bigl(2\op{S}_{z}\op{\rho}\op{S}_{z}-\tfrac{1}{2}\op{\rho}\bigr)                                     \\
                             & +r_{21}\bigl[\op{S}_{-}\op{\rho}\op{S}_{+}-\tfrac{1}{2}(\op{S}_{z}\op{\rho}+\op{\rho}\op{S}_{z}+\op{\rho})\bigr]     \\
                             & +r_{12}\bigl[\op{S}_{+}\op{\rho}\op{S}_{-}+\tfrac{1}{2}(\op{S}_{z}\op{\rho}+\op{\rho}\op{S}_{z}-\op{\rho})\bigr]\, ,
  \end{split}
  \label{eqn-dissipation-superoperator}
  \vspace{-2mm}
\end{equation}
where $r_{12}$ denotes the scattering rate from $\ket{\downarrow}$ to $\ket{\uparrow}$ (vice versa for $r_{21}$) and $r_{\mathrm{p}}$ the pure dephasing rate. From the complete time evolution equation for our system
\vspace{-1.5mm}
\begin{equation}
  \dd{\op{\rho}}{t}=-\frac{\i}{\hbar}\commutator{\op{H}_{\mathrm{A}}+\op{H}_{\mathrm{I},\mathrm{DA}}}{\hat{\rho}}+\mathcal{D}(\hat{\rho})\, ,
  \vspace{-2mm}
\end{equation}
we can derive the familiar full-wave Bloch equations
\vspace{-1.5mm}
\begin{equation}
  \begin{alignedat}{2}
     & \dd{\rho_{21}}{t} &  & =-\i\Omega\rho_{21}-\frac{\i}{\hbar}m_{21}E_{z}\nu-\gamma_{2}\rho_{21}\, ,                                                \\
     & \dd{\nu}{t}       &  & =2\frac{\i}{\hbar}\left(m_{21}E_{z}\rho_{21}^{\ast}-m_{21}^{\ast}E_{z}\rho_{21}\right)-\gamma_{1}\left(\nu-\nu_{0}\right)
    \label{eqn-bloch-equations}
  \end{alignedat}
  \vspace{-1mm}
\end{equation}
for the coherence $\rho_{21}$ and population inversion $\nu=\rho_{22}-\rho_{11}$ (by virtue of the hermiticity $\op{\rho}^{\dagger}=\op{\rho}$ no separate equation for $\rho_{12}=\rho_{21}^{\ast}$ is needed). Here the relaxation rates $\gamma_{1}$, $\gamma_{2}$ and the steady state inversion $\nu_{0}$ are given by
\vspace{-1.5mm}
\begin{equation}
  \begin{alignedat}{2}
     & \gamma_{1} &  & =r_{12}+r_{21}\, ,                                         \\
     & \gamma_{2} &  & =\tfrac{1}{2}\left(r_{12}+r_{21}\right)+r_{\mathrm{p}}\, , \\
     & \nu_{0}    &  & =\frac{r_{12}-r_{21}}{r_{12}+r_{21}}\, .
  \end{alignedat}
  \vspace{-1mm}
\end{equation}
Indeed, the form of \eqref{eqn-dissipation-superoperator} was chosen specifically to yield these results, which are in agreement with an intuitive understanding of the scattering and dephasing process.

Now, the Bloch equations are closed by coupling them to the 1D~Maxwell equations
\vspace{-1.5mm}
\begin{equation}
  \begin{alignedat}{2}
     & \pdd{E_{z}}{x} &  & =\mu_{0}\pdd{H_{y}}{t}\, ,                        \\
     & \pdd{H_{y}}{x} &  & =\varepsilon_{0}\pdd{E_{z}}{t}+\pdd{P_{z}}{t}\, ,
  \end{alignedat}
  \label{eqn-maxwell-equations}
  \vspace{-1mm}
\end{equation}
containing the macroscopic polarization
\vspace{-1.5mm}
\begin{equation}
  P_{z}=n_{\mathrm{3D}}\braket{\op{m}}=2n_{\mathrm{3D}}\Re{m_{21}^{\ast}\rho_{21}}\, .
  \label{eqn-macroscopic-polarization}
  \vspace{-1mm}
\end{equation}
Here $n_{\mathrm{3D}}$ denotes the number density of atoms. The resulting semiclassical MB equations are well suited for optoelectronic device simulations whenever the reduction of the 3D~Maxwell equations to 1D can be justified. For example, semiconductor lasers often possess a waveguide geometry where the cavity cross section remains constant along the optical axis. In this case the transverse mode proﬁle calculation can be decoupled from the longitudinal propagation simulation~\cite{jirauschek_optoelectronic_2019}. The standard finite-difference time-domain  (FDTD) method~\cite{taflove_computational_2005,yee_numerical_1966} for the numerical solution of the 1D~Maxwell equations offers a huge computational benefit compared to the numerical treatment of the quantized optical field from Sec.~\ref{sec-quantization}. Further details concerning the combined numerical solution of the MB equations, including the appropriate treatment of the coupling between them, can be found in~\cite{jirauschek_optoelectronic_2019}.

In conclusion of this section, we note that plugging an ansatz of the form \eqref{eqn-fields-qm} into \eqref{eqn-maxwell-equations} (without the polarization term) yields the conditions
\vspace{-2mm}
\begin{equation}
  \begin{alignedat}{2}
     & \dd{e_{n}}{t} &  & =\omega_{n}h_{n}\, ,  \\
     & \dd{h_{n}}{t} &  & =-\omega_{n}e_{n}\, ,
  \end{alignedat}
  \label{eqn-maxwell-equations-modes-qm}
  \vspace{-2mm}
\end{equation}
which can be regarded as the cavity mode form of the 1D~Maxwell equations. We make use of this observation in the derivation of the stochastic MB equations in Sec.~\ref{sec-variables}.

\section{\label{sec-sde}\!\!From the positive \texorpdfstring{$\bm{P}$}{$P$} representation to SDEs}

Given a quantum mechanical system with Hamiltonian $\op{H}$, our goal is to specify a $c$-number stochastic process that captures the dynamics of the von~Neumann equation
\begin{equation}
  \dd{\op{\rho}}{t}=-\frac{\i}{\hbar}\commutator{\op{H}}{\op{\rho}}\, ,
  \label{eqn-von-neumann-equation}
\end{equation}
for the density operator $\op{\rho}$. To this end, we make the ansatz
\begin{equation}
  \function{\op{\rho}}{t}=\int\function{P}{x,t}\function{\op{\Lambda}}{x}\diff\function{\mu}{x},\quad x=(x_{1},\ldots,x_{n})\, ,
  \label{eqn-integral-representation}
\end{equation}
where $P$ should behave like a probability distribution on a complex nonclassical phase space~\cite{carmichael_statistical_2_2007}. As noted in~\cite{drummond_quantum_1991}, a consistent replacement of operators by $c$-numbers generally requires a nonclassical phase space. Here the integration measure
\begin{equation}
  \diff\function{\mu}{x}=\diff^{2}x_{1}\cdots\diff^{2}x_{n}
\end{equation}
calls for separate integration over all real and imaginary parts. Let us assume that the kernel $\op{\Lambda}$ of this integral representation satisfies
\begin{equation}
  \commutator{\op{H}}{\function{\op{\Lambda}}{x}}=\i\hbar\Biggl[\sum_{i=1}^{n}\function{A_{i}}{x}\pdd{}{x_{i}}+\frac{1}{2}\sum_{i,j=1}^{n}\function{D_{ij}}{x}\frac{\partial^2}{\partial x_{i}\partial x_{j}}\Biggr]\function{\op{\Lambda}}{x}\, ,\label{eqn-commutator-hamiltonian-lambda}
\end{equation}
i.e., only partial derivatives with respect to the phase space variables of order one and two occur. Combining \eqref{eqn-von-neumann-equation}, \eqref{eqn-integral-representation}, and \eqref{eqn-commutator-hamiltonian-lambda} yields
\begin{widetext}
  \vspace{-2.5mm}
  \begin{equation}
    \begin{split}
      \int\function{\op{\Lambda}}{x}\pdd{}{t}\function{P}{x,t}\diff\function{\mu}{x} & =-\frac{\i}{\hbar}\int \function{P}{x,t}\commutator{\op{H}}{\function{\op{\Lambda}}{x}}\diff\function{\mu}{x}                                                                                                                          \\
                                                                                     & =\int \function{P}{x,t}\left(\sum_{i=1}^{n}\function{A_{i}}{x}\pdd{}{x_{i}}+\frac{1}{2}\sum_{i,j=1}^{n}\function{D_{ij}}{x}\frac{\partial^2}{\partial x_{i}\partial x_{j}}\right)\function{\op{\Lambda}}{x}\diff\function{\mu}{x}      \\
                                                                                     & =\int \function{\op{\Lambda}}{x}\left(-\sum_{i=1}^{n}\pdd{}{x_{i}}\function{A_{i}}{x}+\frac{1}{2}\sum_{i,j=1}^{n}\frac{\partial^2}{\partial x_{i}\partial x_{j}}\function{D_{ij}}{x}\right)\function{P}{x,t}\diff\function{\mu}{x}\, ,
    \end{split}
    \label{eqn-fpe-integral-form}
  \end{equation}
  \vspace{-3.5mm}
\end{widetext}
provided that all boundary terms from partial integration vanish. By getting rid of the integrals, we obtain the FPE
\begin{equation}
  \begin{split}
    \pdd{}{t}\function{P}{x,t}=\Biggl[\!-\!\sum_{i=1}^{n}\pdd{}{x_{i}}\function{A_{i}}{x}+\frac{1}{2}\!\sum_{i,j=1}^{n}\frac{\partial^2}{\partial x_{i}\partial x_{j}}\function{D_{ij}}{x}\!\Biggr]\function{P}{x,t}\, ,
  \end{split}
  \label{eqn-fpe}
\end{equation}
with a drift vector $\function{\mat{A}}{x}=\left[\function{A_{i}}{x}\right]\in\Cnums^{n}$ and a diffusion matrix $\function{\mat{D}}{x}=\left[\function{D_{ij}}{x}\right]\in\Cnums^{n\times n}$. It is understood that in the last line of \eqref{eqn-fpe-integral-form} as well as in \eqref{eqn-fpe} the differential operator $\partial/\partial x_{i}$ acts on the whole product $\function{A_{i}}{x}\function{P}{x,t}$ and analogously for the remaining second-order differential operators. If we can find a (nonunique) factorization
\begin{equation}
  \function{\mat{D}}{x}=\function{\mat{B}}{x}\function{\mat{B}}{x}^{\transp}
  \label{eqn-matrix-factorization}
\end{equation}
with a noise matrix $\function{\mat{B}}{x}\in\Cnums^{n\times m}$, then \eqref{eqn-fpe} is equivalent to the It\^o SDE
\begin{equation}
  \diff X_{t} = \function{\mat{A}}{X_{t}}\diff t+\function{\mat{B}}{X_{t}}\diff\mat{W}_{t}\, ,
  \label{eqn-sde}
\end{equation}
driven by the $m$-dimensional Wiener process $\mat{W}$~\cite{carmichael_statistical_1_1999,gardiner_stochastic_2009}. Its solution $X=\set{X(t)=X_{t}\given t\in [0,T]}$ on a time interval $[0,T]$ is the stochastic process we are looking for. Note that we clearly distinguish between stochastic processes and their realizations (e.g., $X$ and $x$) by using uppercase and lowercase letters, respectively.

Before continuing, a few comments are in order: Another way to write the SDE, namely,
\begin{equation}
  \dd{X_{t}}{t} = \function{\mat{A}}{X_{t}}+\function{\mat{B}}{X_{t}}\bm\xi_{t}\, ,
\end{equation}
uses the concept of white noise $\bm\xi=\mathrm{d}\mat{W}/\mathrm{d}t$ and highlights the connection to Langevin forces~\cite{vanKampen_stochastic_2007}. Furthermore, under the necessary assumption of $\mat{D}=\mat{D}^{\transp}$, the existence of a corresponding $\mat{B}$ satisfying \eqref{eqn-matrix-factorization} is guaranteed by the Autonne-Takagi factorization~\cite{horn_matrix_2013}. For numerical reasons, it is important to be able to specify such a factorization explicitly. As exemplified later in this section, the freedom to choose a wide rectangular (i.e., more columns than rows) instead of quadratic $\mat{B}$ can prove useful here, but the additional noise dimensions are also a numerical drawback by themselves. Regarding the attainable simulation time, numerical diffusion gauges realized by algorithmically solving for $\mat{B}$ in each time step can even be beneficial~\cite{rousse_simulations_2024}. In general, the inherent nonuniqueness of the diffusion matrix factorization can always be interpreted as a diffusion gauge~\cite{deuar_gauge_2002}, which manifests itself as a degree of freedom in the statement of the SDE but does not affect its solution, i.e., the stochastic process.

\subsection{\label{sec-projectors}Bosonic and fermionic projectors}

We want to apply the procedure above to the full-wave Jaynes-Cummings-type system described in Sec.~\ref{sec-quantization}, culminating in an SDE. For that, we first need to construct an appropriate kernel (see Sec.~\ref{sec-jaynes-cummings}). Owing to the mixed bosonic and fermionic nature of the involved Hamiltonian, the preparatory work is done in two separate but related steps.

The case of purely bosonic systems is well known: For the bosonic projector
\vspace{-1.5mm}
\begin{equation}
  \op{\Lambda}_{\mathrm{F}}(\alpha,\beta)=\frac{\ket{\alpha}\bra{\beta^{\ast}}}{\braket{\beta^{\ast}|\alpha}}
  \label{eqn-bosonic-lambda}
  \vspace{-2mm}
\end{equation}
onto the normalized coherent states
\vspace{-1.5mm}
\begin{equation}
  \ket{\alpha}=\e^{-\frac{\abs{\alpha}^2}{2}}\exp(\alpha{\op{a}}^{\dagger})\ket{0}\, ,
  \label{eqn-coherent-state}
  \vspace{-2mm}
\end{equation}
our ansatz \eqref{eqn-integral-representation} corresponds to the positive $P$ representation, which has proven to be very useful in quantum optics~\cite{drummond_generalised_1980,kheruntsyan_einstein_2005,werner_phase_1997,wolinsky_quantum_1988}. The off-diagonal projectors double the phase space dimension [i.e., $(\alpha,\beta)$ replaces $(\alpha,\alpha^{\ast})$], and these additional dimensions are known to capture nonclassical light features~\cite{walls_quantum_1994}. Because of the projector differential identities
\vspace{-1.5mm}
\begin{equation}
  \begin{alignedat}{2}
     & \op{a}^{\dagger}\op{\Lambda}_{\mathrm{F}} &  & =\left(\pdd{}{\alpha}+\beta\right)\op{\Lambda}_{\mathrm{F}}\, , \\
     & \op{a}\op{\Lambda}_{\mathrm{F}}           &  & =\alpha\op{\Lambda}_{\mathrm{F}}\, ,                            \\
     & \op{\Lambda}_{\mathrm{F}}\op{a}^{\dagger} &  & =\beta\op{\Lambda}_{\mathrm{F}}\, ,                             \\
     & \op{\Lambda}_{\mathrm{F}}\op{a}           &  & =\left(\pdd{}{\beta}+\alpha\right)\op{\Lambda}_{\mathrm{F}}\, ,
  \end{alignedat}
  \label{eqn-bosonic-formulas}
  \vspace{-2mm}
\end{equation}
we get an expression of the desired form \eqref{eqn-commutator-hamiltonian-lambda} for suitable Hamiltonians. In this context, it is also important that there is an explicit formula for an initial probability distribution in terms of a given initial density operator, even though such a probability distribution need not be unique.

Next, we need to find a way to also deal with the fermionic component of a given system. An existing approach~\cite{barry_qubit_2008,mandt_stochastic_2015,ng_simulation_2013} is based on coherent spin states \cite{radcliffe_properties_1971}. For spin~$s=1/2$ and $\ket{s,s}=\ket{\uparrow}$ their definition boils down to
\vspace{-1.5mm}
\begin{equation}
  \begin{split}
    \ket{z} & =\left(1+\abs{z}^{2}\right)^{-s}\exp(z\op{S}_{-})\ket{s,s}                      \\
            & =\frac{1}{\sqrt{1+\abs{z}^{2}}}\bigl(z\ket{\downarrow}+\ket{\uparrow}\bigr)\, .
  \end{split}
  \label{eqn-coherent-spin-states}
  \vspace{-2mm}
\end{equation}
This motivates the more general ansatz for nonorthogonal fermionic states
\vspace{-1.5mm}
\begin{equation}
  \ket{|z}=\function{f}{z}\ket{\downarrow}+\function{g}{z}\ket{\uparrow}\, ,
  \label{eqn-analytic-states}
  \vspace{-2mm}
\end{equation}
with analytic coefficient functions $f$ and $g$. The notation $\ket{|\cdot}$ (see~\cite{deuar_gauge_2002}) is chosen to remind us of the fact that these states might be unnormalized. Not only will the resulting freedom to fine-tune $f$ and $g$ pay off later, but the extra generality is also helpful for separating inherent features of the problem from details due to a specific choice of these analytic coefficient functions. However, an obvious drawback of renouncing the coherent spin states is that an extension to the case of an $n$-level atom is not straightforward. It is clear from the local power series expansions
\vspace{-1.5mm}
\begin{equation}
  \begin{split}
    \function{f}{z} & =\sum_{k=0}^{\infty}f_{k}(z-z_{0})^{k}\, , \\
    \function{g}{z} & =\sum_{k=0}^{\infty}g_{k}(z-z_{0})^{k}\, ,
  \end{split}
  \label{eqn-power-series}
  \vspace{-2mm}
\end{equation}
that $\tilde{f}$ and $\tilde{g}$ defined by
\vspace{-1.5mm}
\begin{equation}
  \begin{split}
    \function{\tilde{f}}{z} & =\function{f}{z^{\ast}}^{\ast}=\sum_{k=0}^{\infty}f_{k}^{\ast}(z-{z_{0}}^{\ast})^{k}\, , \\
    \function{\tilde{g}}{z} & =\function{g}{z^{\ast}}^{\ast}=\sum_{k=0}^{\infty}g_{k}^{\ast}(z-{z_{0}}^{\ast})^{k}
  \end{split}
  \label{eqn-auxiliary-functions}
  \vspace{-2mm}
\end{equation}
are also analytic functions. Moreover $h=g/f$ and $\tilde{h}=\tilde{g}/\tilde{f}$ are analytic where they are defined. Now, we define the fermionic projector
\vspace{-1.5mm}
\begin{equation}
  \begin{split}
    \function{\op{\Lambda}_{\mathrm{A}}}{z,w}= & \,\frac{\ket{|z}\bra{w^{\ast}|}}{\braket{w^{\ast}\lvert\rvert z}}                                                                              \\
    =                                          & \,\frac{1}{1+\function{h}{z}\function{\tilde{h}}{w}}\bigl[\ketbra{\downarrow}{\downarrow}+\function{\tilde{h}}{w}\ketbra{\downarrow}{\uparrow} \\
                                               & +\function{h}{z}\ketbra{\uparrow}{\downarrow}+\function{h}{z}\function{\tilde{h}}{w}\ketbra{\uparrow}{\uparrow}\bigr],
  \end{split}
  \label{eqn-fermionic-lambda}
  \vspace{-2mm}
\end{equation}
in analogy to \eqref{eqn-bosonic-lambda}. It can be shown by a lengthy but direct calculation that the projector differential identities
\vspace{-1.5mm}
\begingroup
\allowdisplaybreaks
\begin{align}
  \op{S}_{+}\op{\Lambda}_{\mathrm{A}} & =\left[\frac{1}{\function{h'}{z}}\pdd{}{z}+\frac{\function{\tilde{h}}{w}}{1+\function{h}{z}\function{\tilde{h}}{w}}\right]\op{\Lambda}_{\mathrm{A}}\, ,\nonumber                                                   \\
  \op{S}_{-}\op{\Lambda}_{\mathrm{A}} & =\left[-\frac{\function{h}{z}^{2}}{\function{h'}{z}}\pdd{}{z}+\frac{\function{h}{z}}{1+\function{h}{z}\function{\tilde{h}}{w}}\right]\op{\Lambda}_{\mathrm{A}}\, ,\nonumber                                        \\
  \op{S}_{z}\op{\Lambda}_{\mathrm{A}} & =\left\{\frac{\function{h}{z}}{\function{h'}{z}}\pdd{}{z}-\frac{1-\function{h}{z}\function{\tilde{h}}{w}}{2\left[1+\function{h}{z}\function{\tilde{h}}{w}\right]}\right\}\op{\Lambda}_{\mathrm{A}}\, ,\nonumber    \\
  \op{\Lambda}_{\mathrm{A}}\op{S}_{+} & =\left[-\frac{\function{\tilde{h}}{w}^{2}}{\function{\tilde{h}'}{w}}\pdd{}{w}+\frac{\function{\tilde{h}}{w}}{1+\function{h}{z}\function{\tilde{h}}{w}}\right]\op{\Lambda}_{\mathrm{A}}\, ,\nonumber                \\
  \op{\Lambda}_{\mathrm{A}}\op{S}_{-} & =\left[\frac{1}{\function{\tilde{h}'}{w}}\pdd{}{w}+\frac{\function{h}{z}}{1+\function{h}{z}\function{\tilde{h}}{w}}\right]\op{\Lambda}_{\mathrm{A}}\, ,\nonumber                                                   \\
  \op{\Lambda}_{\mathrm{A}}\op{S}_{z} & =\left\{\frac{\function{\tilde{h}}{w}}{\function{\tilde{h}'}{w}}\pdd{}{w}-\frac{1-\function{h}{z}\function{\tilde{h}}{w}}{2\left[1+\function{h}{z}\function{\tilde{h}}{w}\right]}\right\}\op{\Lambda}_{\mathrm{A}}
  \label{eqn-fermionic-formulas}
  \vspace{-2mm}
\end{align}

\endgroup
hold. This is a generalization and unification of some of the projector differential identities arising from coherent spin states that can be found in~\cite{mandt_stochastic_2015,ng_simulation_2013}. As already discussed for the bosonic case, it is essential to specify an explicit way to write an initial atomic density operator, in the form of \eqref{eqn-integral-representation}, where $P$ is a probability distribution. Otherwise, the states \eqref{eqn-analytic-states} would turn out to be unusable since it is not sufficient to state an SDE; one also has to be able to supply it with initial values. The details are quite technical and can be found in Appendix~\ref{sec-initialization}. Compared to their counterpart \eqref{eqn-bosonic-formulas}, the projector differential identities \eqref{eqn-fermionic-formulas} are highly nonlinear and have singularities. Another positive $P$ representation approach~\cite{chuchurka_quantum_2023,ng_real-time_2011,olsen_phase-space_2005}, which uses Schwinger bosons~\cite{sakurai_modern_1995} for dealing with the fermionic system component, can mitigate this problem at the expense of much broader and therefore harder to sample initial probability distributions.
\vspace{-2mm}

\subsection{\label{sec-jaynes-cummings}The full-wave Jaynes-Cummings-type SDE}

Now we are ready to focus on the full-wave Jaynes-Cummings-type system. For the rest of the discussion, we restrict ourselves to a finite number $N\geq1$ of cavity modes in the Hamiltonian \eqref{eqn-hamiltonian-jc} since we are not equipped to deal with an infinite dimensional phase space. It is natural to consider the phase space variables
\begin{equation}
  \phi=(\phi_{1},\ldots,\phi_{2(N+1)})=(\alpha_{1},\beta_{1},\ldots,\alpha_{N},\beta_{N},z,w)\, ,
\end{equation}
which belong to the kernel
\begin{equation}
  \function{\op{\Lambda}_{\mathrm{JC}}}{\phi}=\function{\op{\Lambda}_{\mathrm{F}}}{\alpha_{1},\beta_{1}}\otimes\cdots\otimes\function{\op{\Lambda}_{\mathrm{F}}}{\alpha_{N},\beta_{N}}\otimes\function{\op{\Lambda}_{\mathrm{A}}}{z,w}\, ,
  \label{eqn-lambda-jc}
\end{equation}
built from the bosonic and fermionic projectors \eqref{eqn-bosonic-lambda} and \eqref{eqn-fermionic-lambda}, respectively. Evaluating $\commutator{\op{H}_{\mathrm{JC}}}{\op{\Lambda}_{\mathrm{JC}}}$ with the help of the projector differential identities \eqref{eqn-bosonic-formulas} and \eqref{eqn-fermionic-formulas} leads to a FPE with drift vector
\begin{equation}
  \function{\mat{A}_{\mathrm{JC}}}{\phi}=\i\scalebox{0.88}{$\displaystyle\begin{bmatrix}
        -\omega_{1}\alpha_{1}-\function{g}{\omega_{1}}\sin(k_{1}x_{0})\frac{\function{h}{z}+\function{\tilde{h}}{w}}{1+\function{h}{z}\function{\tilde{h}}{w}}                                \\
        \omega_{1}\beta_{1}+\function{g}{\omega_{1}}\sin(k_{1}x_{0})\frac{\function{h}{z}+\function{\tilde{h}}{w}}{1+\function{h}{z}\function{\tilde{h}}{w}}                                  \\
        \vdots                                                                                                                                                                                \\
        -\omega_{N}\alpha_{N}-\function{g}{\omega_{N}}\sin(k_{N}x_{0})\frac{\function{h}{z}+\function{\tilde{h}}{w}}{1+\function{h}{z}\function{\tilde{h}}{w}}                                \\
        \omega_{N}\beta_{N}+\function{g}{\omega_{N}}\sin(k_{N}x_{0})\frac{\function{h}{z}+\function{\tilde{h}}{w}}{1+\function{h}{z}\function{\tilde{h}}{w}}                                  \\
        -\Omega\frac{\function{h}{z}}{\function{h'}{z}}+\sum_{n=1}^{N}\function{g}{\omega_{n}}\left(\alpha_{n}+\beta_{n}\right)\sin(k_{n}x_{0})\frac{\function{h}{z}^{2}-1}{\function{h'}{z}} \\
        \Omega\frac{\function{\tilde{h}}{w}}{\function{\tilde{h}'}{w}}-\sum_{n=1}^{N}\function{g}{\omega_{n}}\left(\alpha_{n}+\beta_{n}\right)\sin(k_{n}x_{0})\frac{\function{\tilde{h}}{w}^{2}-1}{\function{\tilde{h}'}{w}}
      \end{bmatrix}$}\, .\label{eqn-drift-vector-jc}
\end{equation}
The diffusion matrix
\begin{equation}
  \function{\mat{D}_{\mathrm{JC}}}{\phi}=\i\begin{bmatrix}
    \mat{0}                      & \cdots & \mat{0}                      & \function{\mat{D}_{1}}{\phi} \\
    \vdots                       & \ddots & \vdots                       & \vdots                       \\
    \mat{0}                      & \cdots & \mat{0}                      & \function{\mat{D}_{N}}{\phi} \\
    \function{\mat{D}_{1}}{\phi} & \cdots & \function{\mat{D}_{N}}{\phi} & \mat{0}
  \end{bmatrix}
  \label{eqn-diffusion-matrix-jc}
\end{equation}
is built from $2\times2$ blocks. The nonzero blocks
\vspace{-0.5mm}
\begin{equation}
  \function{\mat{D}_{n}}{\phi}=\begin{bmatrix}
    \function{d_{n}}{\phi} & 0                               \\
    0                      & -\function{\tilde{d}_{n}}{\phi}
  \end{bmatrix}
\end{equation}
contain the entries
\vspace{-0.5mm}
\begin{equation}
  \begin{split}
    \function{d_{n}}{\phi}         & =\function{g}{\omega_{n}}\sin(k_{n}x_{0})\frac{\function{h}{z}^{2}-1}{\function{h'}{z}}\, ,                 \\
    \function{\tilde{d}_{n}}{\phi} & =\function{g}{\omega_{n}}\sin(k_{n}x_{0})\frac{\function{\tilde{h}}{w}^{2}-1}{\function{\tilde{h}'}{w}}\, ,
  \end{split}
  \label{eqn-diffusion-matrix-jc-entries}
\end{equation}
which depend on the cavity mode number, the bosonic-fermionic coupling constants, the position of the TLA in the cavity, and the fermionic phase space variables. In this way, we have identified all factors contributing to the inherent quantum noise. Before we can state the desired SDE for the system, we need to factorize $\mat{D}_{\mathrm{JC}}$. To this end consider
\vspace{-1mm}
\begin{equation}
  \begin{alignedat}{2}
     & \function{\mat{P}_{n}}{\phi} &  & =\sqrt{\frac{\function{d_{n}}{\phi}}{2}}\begin{bmatrix}
                                                                                   \i & -1 \\
                                                                                   0  & 0
                                                                                 \end{bmatrix}\, ,         \\
     & \function{\mat{Q}_{n}}{\phi} &  & =\sqrt{\frac{\function{\tilde{d}_{n}}{\phi}}{2}}\begin{bmatrix}
                                                                                           0   & 0  \\
                                                                                           -\i & -1
                                                                                         \end{bmatrix}\, , \\
     & \function{\mat{R}_{n}}{\phi} &  & =\sqrt{\frac{\function{d_{n}}{\phi}}{2}}\begin{bmatrix}
                                                                                   -\i & -1 \\
                                                                                   0   & 0
                                                                                 \end{bmatrix}\, ,         \\
     & \function{\mat{S}_{n}}{\phi} &  & =\sqrt{\frac{\function{\tilde{d}_{n}}{\phi}}{2}}\begin{bmatrix}
                                                                                           0   & 0 \\
                                                                                           -\i & 1
                                                                                         \end{bmatrix}\, ,
  \end{alignedat}
\end{equation}
and set
\vspace{-1mm}
\begin{equation}
  \function{\mat{B}_{n}}{\phi}=\begin{bmatrix}
    \mat{0} & \cdots & \mat{0} & \mat{0}                      & \mat{0} & \cdots & \mat{0} & \mat{0}                      \\
    \vdots  & \ddots & \vdots  & \vdots                       & \vdots  & \ddots & \vdots  & \vdots                       \\
    \mat{0} & \cdots & \mat{0} & \mat{0}                      & \mat{0} & \cdots & \mat{0} & \mat{0}                      \\
    \mat{0} & \cdots & \mat{0} & \function{\mat{P}_{n}}{\phi} & \mat{0} & \cdots & \mat{0} & \function{\mat{Q}_{n}}{\phi} \\
    \mat{0} & \cdots & \mat{0} & \mat{0}                      & \mat{0} & \cdots & \mat{0} & \mat{0}                      \\
    \vdots  & \ddots & \vdots  & \vdots                       & \vdots  & \ddots & \vdots  & \vdots                       \\
    \mat{0} & \cdots & \mat{0} & \mat{0}                      & \mat{0} & \cdots & \mat{0} & \mat{0}                      \\
    \mat{0} & \cdots & \mat{0} & \function{\mat{R}_{n}}{\phi} & \mat{0} & \cdots & \mat{0} & \function{\mat{S}_{n}}{\phi}
  \end{bmatrix}\, ,
  \label{eqn-bn}
\end{equation}
where the nonzero blocks are located in row $n$ or $N+1$ and column $n$ or $N+1$ of the block matrix. Then it is straightforward to check that
\vspace{-0.5mm}
\begin{equation}
  \function{\mat{B}_{\mathrm{JC}}}{\phi}=\sqrt{\i}\begin{bmatrix}
    \function{\mat{B}_{1}}{\phi} & \cdots & \function{\mat{B}_{N}}{\phi}
  \end{bmatrix}
  \label{eqn-noise-matrix-jc}
\end{equation}
satisfies the requirement $\mat{B}_{\mathrm{JC}}\mat{B}_{\mathrm{JC}}^{\transp}=\mat{D}_{\mathrm{JC}}$. This construction is inspired by~\cite{mandt_stochastic_2015} where the structure of \eqref{eqn-bn} can be found for the special case $n=N=1$, RWA, and $\function{h}{z}=z$ (implying $\function{\tilde{h}}{w}=w$), which corresponds to the coherent spin states \eqref{eqn-coherent-spin-states}. It would be an improvement to find a smarter factorization that requires fewer noise dimensions. The resulting SDE
\vspace{-0.5mm}
\begin{equation}
  \diff \Phi_{t}=\function{\mat{A}_{\mathrm{JC}}}{\Phi_{t}}\diff t+\function{\mat{B}_{\mathrm{JC}}}{\Phi_{t}}\diff\mat{W}_{t}
  \label{eqn-sde-jc}
\end{equation}
is nonscalar and nonlinear, and generally has nonadditive noise, i.e., the noise matrix \eqref{eqn-noise-matrix-jc} is not constant. Therefore, searching for analytic solutions is probably out of the question, and even numerical solution methods can struggle.

\noindent Surprisingly, we are able to address the nonadditive noise issue: For the parameter $\delta\in\Cnums$ the ordinary differential equation (ODE)
\vspace{-2mm}
\begin{equation}
  \function{h'}{z}=\frac{1}{\delta}\left[\function{h}{z}^{2}-1\right]
  \vspace{-1.5mm}
\end{equation}
has the family of solutions
\vspace{-1.5mm}
\begin{equation}
  \function{h}{z}=\frac{1-\exp\bigl(\frac{2z}{\delta}+\kappa\bigr)}{1+\exp\bigl(\frac{2z}{\delta}+\kappa\bigr)},\quad\kappa\in\Cnums\, .
  \vspace{-2mm}
\end{equation}
Here the parameter $\delta$ is a degree of freedom directly linked to the diffusion via \eqref{eqn-diffusion-matrix-jc-entries}. Consequently, the new full-wave additive noise states with, e.g., $\kappa=0$
\vspace{-1.5mm}
\begin{equation}
  \begin{split}
    \ket{|z}=\bigl(1+\e^{\frac{2z}{\delta}}\bigr)\ket{\downarrow}+\bigl(1-\e^{\frac{2z}{\delta}}\bigr)\ket{\uparrow}\, ,
  \end{split}
  \label{eqn-exponential-states}
  \vspace{-2mm}
\end{equation}
have the special property that the diffusion matrix \eqref{eqn-diffusion-matrix-jc} and the noise matrix \eqref{eqn-noise-matrix-jc} are constant. We note that this kind of optimization, which can also apply to other system Hamiltonians, is not achievable with the well-known diffusion gauges (see Sec.~\ref{sec-sde}). It is an interesting fact that the coherent spin states \eqref{eqn-coherent-spin-states} themselves lead to additive noise for the SDE associated with the RWA Jaynes-Cummings model. The benefit can lie in the order of convergence of a numerical solution method: For example, the basic Euler-Maruyama method (weak order $\Delta t$, strong order $\sqrt{\Delta t}$) is equivalent to the Milstein method (weak order $\Delta t$, strong order $\Delta t$) for the case of additive noise~\cite{kloeden_stochastic_1992}. Moreover, it is plausible that the stability of numerical solution methods can improve as well. A discussion of the role of additive versus nonadditive noise in physics can be found in~\cite{schenzle_multiplicative_1979}.
\vspace{-3mm}

\section{\label{sec-simulation}SDE simulation challenges}

Even though SDE simulations are not the focus of this paper, we want to present a simple example for three reasons: to test the results of Sec.~\ref{sec-sde} while demonstrating the desirability of additive noise in particular, to explain some aspects of SDE postprocessing that are relevant for Sec.~\ref{sec-variables}, and to recapitulate some general problems that affect SDE based approaches (see Sec.~\ref{sec-conclusion}).

The big advantage of SDEs derived by means of a nonorthogonal basis state expansion is that the sizes of the associated drift vector and noise matrix scale linearly with the complexity of the system Hamiltonian, as opposed to the exponential complexity scaling for a straightforward orthogonal basis state expansion~\cite{barry_qubit_2008,ng_real-time_2011}. Contrary to this observation, we reduce the complexity by considering only one cavity mode in the Hamiltonian \eqref{eqn-hamiltonian-jc}, which yields the full-wave Jaynes-Cummings model. This does not play to the strengths of SDEs, but it makes it easy to compare the results of the SDE simulation with those of an independent Heisenberg picture reference simulation resorting to a truncated number state basis for the representation of the operators by matrices. We note that before the semiclassical limit is eventually reached, the high photon number cutoff for the truncated number state basis of each bosonic mode causes trouble for this Hilbert space approach. Due to its $c$-number nature, the SDE is not negatively affected at all.

Unless otherwise indicated, we use the full-wave additive noise states \eqref{eqn-exponential-states} with parameter $\delta=4$. Then we obtain the drift vector
\vspace{-1mm}
\begin{equation}
  \function{\mat{A}_{\mathrm{JC}}}{\phi}=\i\begin{bmatrix}
    -\omega\alpha+gs\tanh\left(\frac{z}{\delta}+\frac{w}{\delta^{\ast}}\right)                   \\
    \omega\beta-gs\tanh\left(\frac{z}{\delta}+\frac{w}{\delta^{\ast}}\right)                     \\
    -\frac{\Omega\delta}{2}\sinh\left(\frac{2z}{\delta}\right)+gs\delta\left(\alpha+\beta\right) \\
    \frac{\Omega\delta^{\ast}}{2}\sinh\left(\frac{2w}{\delta^{\ast}}\right)-gs\delta^{\ast}\left(\alpha+\beta\right)
  \end{bmatrix}
  \label{eqn-drift-vector-jc-special}
  \vspace{-1mm}
\end{equation}
and noise matrix
\vspace{-1mm}
\begin{equation}
  \function{\mat{B}_{\mathrm{JC}}}{\phi}=\sqrt{\frac{\i gs}{2}}\begin{bmatrix}
    \i\sqrt{\delta}  & -\sqrt{\delta} & 0                       & 0                     \\
    0                & 0              & -\i\sqrt{\delta^{\ast}} & -\sqrt{\delta^{\ast}} \\
    -\i\sqrt{\delta} & -\sqrt{\delta} & 0                       & 0                     \\
    0                & 0              & -\i\sqrt{\delta^{\ast}} & \sqrt{\delta^{\ast}}
  \end{bmatrix}
  \label{eqn-noise-matrix-jc-special}
  \vspace{-1mm}
\end{equation}
[see \eqref{eqn-drift-vector-jc} and \eqref{eqn-noise-matrix-jc}], where we have set $\omega=\omega_{1}$, $k=k_{1}$, $\alpha=\alpha_{1}$, $\beta=\beta_{1}$, $g=\function{g}{\omega_{1}}$, and $s=\sin(k_{1}x_{0})$, in order to condense the expressions. We observe that $\lim_{\delta\to0}\mat{B}_{\mathrm{JC}}=0$, but the naive idea -- the less noise, the better -- is a bit treacherous. In this context, it would be interesting to investigate whether in the limit $\delta\rightarrow0$ chaos in the drift ODE
\vspace{-1mm}
\begin{equation}
  \dd{\phi}{t}=\function{\mat{A}_{\mathrm{JC}}}{\phi}
  \vspace{-1mm}
\end{equation}
counterbalances the vanishing noise. This question exceeds the scope of the present work but, e.g., \cite{kirilyuk_mechanism_2000} deals with chaos in the MB equations.
\begin{figure}[t]
  \centering
  \tikzexternalenable
  \begin{tikzpicture}
    \begin{groupplot}[
        group style={
            group size=1 by 2,
            ylabels at=edge left,
            xlabels at=edge bottom,
          },
        width=0.45\textwidth,
        height=0.33\textwidth,
        y label style={at={(axis description cs:-0.1,.5)}, anchor=south},
        ylabel={Im},
      ]
      \nextgroupplot[
        legend pos=north east,
        legend cell align=left,
        legend style={nodes={scale=0.8,transform shape}},
        xmin=-0.5,
        xmax=0.5,
        ymin=-0.6,
        ymax=0.6,
        unit markings=parenthesis,
        title = {(a)},
        title style={at={(0.025,0.925)}, anchor=north west},
        xlabel={Re},
        grid=both,
        grid style={draw=TUMGrayLight,dotted,ultra thin},
      ]
      \addplot[color=TUMBlue, thick, solid] table [y=imag_3, x=real_3, col sep=comma] {fig-coherent-state.csv};
      \addplot[color=TUMOrange, thick, solid] table [y=imag_4, x=real_4, col sep=comma] {fig-coherent-state.csv};
      \legend{$z$,$w$}

      \nextgroupplot[
        legend pos=north east,
        legend cell align=left,
        legend style={nodes={scale=0.8,transform shape}},
        xmin=-0.5,
        xmax=2.5,
        ymin=-5,
        ymax=5,
        unit markings = parenthesis,
        title = {(b)},
        title style={at={(0.025,0.925)}, anchor=north west},
        xlabel={Re},
        grid=both,
        grid style={draw=TUMGrayLight,dotted,ultra thin},
      ]
      \addplot[color=TUMBlue, thick, solid] table [y=imag_3, x=real_3, col sep=comma] {fig-exponential-state.csv};
      \addplot[color=TUMOrange, thick, solid] table [y=imag_4, x=real_4, col sep=comma] {fig-exponential-state.csv};
      \legend{$z$,$w$}
    \end{groupplot}
  \end{tikzpicture}
  \tikzexternaldisable
  \caption{Trajectory (for the same time interval) of the fermionic phase space variables $(z,w)$ in the complex plane averaged over 1000 runs of the full-wave Jaynes-Cummings SDE for (a) coherent spin states and (b) full-wave additive noise states. The blue and orange lines belong to $z$ and $w$, respectively.}
  \label{fig-coherent-state-vs-exponential-state}
\end{figure}
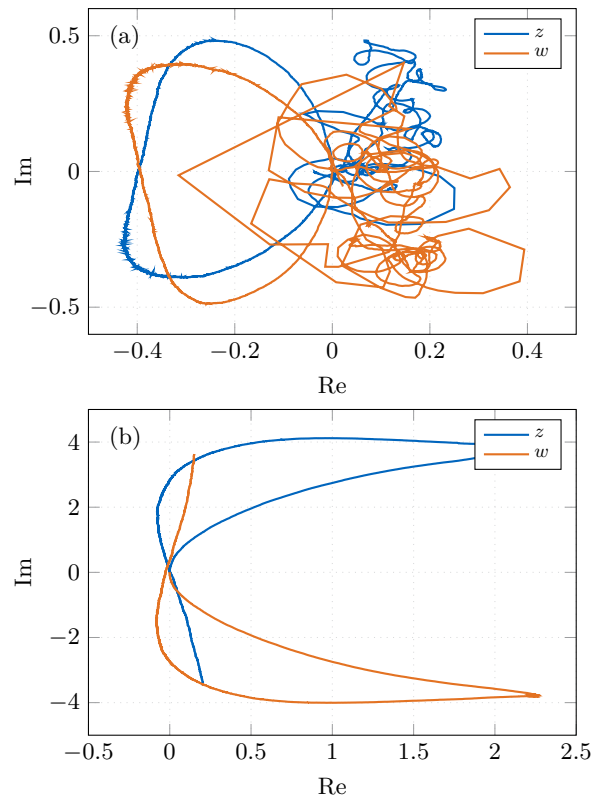
Now we specify the setup of the full-wave Jaynes-Cummings SDE simulation. Let the TLA be positioned in the center $x_{0}=l/2$ of the cavity. Since our interest lies in the qualitative behavior of the SDE determined by \eqref{eqn-drift-vector-jc-special} and \eqref{eqn-noise-matrix-jc-special}, we consider the set of parameters $\hbar=1$, $\Omega=1000$, $\omega=1100$ and $g=200$ with arbitrary units. The initial density operator is taken to be of the form $\op{\rho}_{0}=\op{\rho}_{\mathrm{F},0}\otimes\op{\rho}_{\mathrm{A},0}$ with $\op{\rho}_{\mathrm{F},0}=\ketbra{\alpha}{\alpha}$ and
\vspace{-2mm}
\begin{equation}
  \op{\rho}_{\mathrm{A},0}=\frac{1}{1+\e^{-\beta\hbar\Omega}}\ketbra{\downarrow}{\downarrow}+\frac{1}{1+\e^{\beta\hbar\Omega}}\ketbra{\uparrow}{\uparrow}\, .
  \vspace{-2mm}
\end{equation}
Here the coherence parameter is $\alpha=5$, i.e., there are on average $\abs{\alpha}^{2}=25$ photons in the cavity and the thermal parameter is $\beta=1/(\hbar\Omega)$, i.e., the TLA is inverted with probability $1/(1+\e)\approx0.27$. The initialization of the bosonic phase space variables poses no problems. As mentioned in Sec.~\ref{sec-projectors}, Appendix~\ref{sec-initialization} contains all that is necessary to convert $\op{\rho}_{\mathrm{A},0}$ into initial values for the fermionic phase space variables. For reasons of speed and simplicity, we choose the basic Euler-Maruyama method on the equidistant temporal grid $t_{n}$, $n=1,\ldots,N=\num{8193}$ with $t_{1}=0$ and $t_{N}=\pi/\omega$ (i.e.\ $t_{n+1}-t_{n}=\Delta t\approx3.5\cdot10^{-7}$).

From $R$ independent SDE runs we obtain independent realization paths $\phi_{n}^{r}$ with $r=1,\ldots,R$ and $n=1,\ldots,N$ for the phase space variables. Figure~\ref{fig-coherent-state-vs-exponential-state} demonstrates the impact of the choice of nonorthogonal fermionic basis states on our ability to adequately sample the SDE statistics. For the coherent spin states \eqref{eqn-coherent-spin-states} in \figref[(a)]{fig-coherent-state-vs-exponential-state} \num{1000} SDE runs quickly become insufficient to resolve the statistical expectation values of the fermionic phase space variables $(z,w)$ in the time interval $[0,t_{N}]$. In contrast, \num{1000} SDE runs suffice for all of $[0,t_{N}]$ when we instead use the full-wave additive noise states \eqref{eqn-exponential-states} by virtue of an overall slower diffusion [see \figref[(b)]{fig-coherent-state-vs-exponential-state}]. The fact that more and more SDE runs are needed to resolve the time evolution for longer time scales constitutes a general disadvantage. However, those SDE runs are independent of each other and can thus be parallelized. Figure~\ref{fig-coherent-state-vs-exponential-state} also illustrates that because of the phase space dimension doubling, $z$ and $w$ are complex conjugates only on average, and the same does not apply for individual (or averages of too few) realizations. Most of the time, we are not directly interested in these phase space variables because they might lack a direct physical interpretation. Instead, given an operator, e.g., $\op{S}_{-}$, we ultimately want to compute associated independent realization paths $\sigma_{n}^{r}$ such that
\vspace{-2mm}
\begin{equation}
  \braket{\langle\sigma\rangle}_{n}=\frac{1}{R}\sum_{r=1}^{R}\sigma_{n}^{r}\to\function{\braket{\op{S}_{-}}}{t_{n}}=\trace\left[\op{S}_{-}\function{\op{\rho}}{t_{n}}\right]\, ,
  \label{eqn-operator-ensemble-mean}
  \vspace{-2mm}
\end{equation}
in the limit $R\to\infty$. Here we choose the notation $\braket{\langle\cdot\rangle}$ for the statistical expectation value to distinguish it from the quantum mechanical expectation value. The required postprocessing of the data generated by the SDE runs is described below. Because of \eqref{eqn-bosonic-lambda} and the completeness of the number states, it is easy to see that
\vspace{-2mm}
\begin{equation}
  \sum_{n=0}^{\infty}\braket{n|\function{\op{\Lambda}_{\mathrm{F}}}{\alpha,\beta}|n}=1\, .
\end{equation}
After introducing the function
\vspace{-2mm}
\begin{equation}
  \sigma=\function{v}{\phi}=\frac{\function{h}{z}}{1+\function{h}{z}\function{\tilde{h}}{w}}\, ,
  \label{eqn-expression-s-minus}
  \vspace{-2mm}
\end{equation}
we can write
\vspace{-2mm}
\begin{equation}
  \begin{alignedat}{2}
     & \braket{\downarrow|\op{S}_{-}\function{\op{\Lambda}_{\mathrm{A}}}{z,w}|\downarrow} &  & =v(\phi)\, , \\
     & \braket{\uparrow|\op{S}_{-}\function{\op{\Lambda}_{\mathrm{A}}}{z,w}|\uparrow}     &  & =0
  \end{alignedat}
  \vspace{-2mm}
\end{equation}
by using \eqref{eqn-fermionic-lambda}. Plugging in \eqref{eqn-integral-representation} and simplifying with the help of these results yields
\vspace{-2mm}
\begin{align}
  \trace\left[\op{S}_{-}\function{\op{\rho}}{t}\right] & =\sum_{n=0}^{\infty}\left[\bra{n,\downarrow}\op{S}_{-}\function{\op{\rho}}{t}\ket{n,\downarrow}+\bra{n,\uparrow}\op{S}_{-}\function{\op{\rho}}{t}\ket{n,\uparrow}\right]\nonumber \\
                                                       & =\int\function{P}{\phi,t}\function{v}{\phi}\diff\function{\mu}{\phi}\, .
  \vspace{-2mm}
\end{align}
Hence, it is possible to just set $\sigma_{n}^{r}=\function{v}{\phi_{n}^{r}}$ in \eqref{eqn-operator-ensemble-mean} because due to the equivalence of the SDE and FPE, realizations $\phi_{n}^{r}$ come to lie in the neighborhood $B_{\Delta}$ with the relative frequency $\int_{B_{\Delta}}\function{P}{\phi,t_{n}}\diff\function{\mu}{\phi}$. Although this method for computing $\sigma_{n}^{r}$ is sufficient for most applications, we now discuss a way to get better accuracy at the expense of increased computational load. The derivative of the stochastic process $\Sigma=\function{v}{\Phi}$ is
\begin{widetext}
  \vspace{-3mm}
  \begin{equation}
    \diff\Sigma_{t}=\left[\sum_{i=1}^{n}\function{A_{\mathrm{JC},i}}{\Phi_{t}}\pdd{v}{\phi_{i}}(\Phi_{t})+\frac{1}{2}\sum_{j=1}^{m}\sum_{p,q=1}^{n}\function{B_{\mathrm{JC},pj}}{\Phi_{t}}\function{B_{\mathrm{JC},qj}}{\Phi_{t}}\frac{\partial^{2}v}{\partial\phi_{p}\partial\phi_{q}}(\Phi_{t})\right]\diff t+\sum_{j=1}^{m}\sum_{i=1}^{n}\function{B_{\mathrm{JC},ij}}{\Phi_{t}}\pdd{v}{\phi_{i}}(\Phi_{t})\diff W_{j,t}\label{eqn-ito-formula}
    \vspace{-3mm}
  \end{equation}
\end{widetext}
by It\^o's formula~\cite{kloeden_stochastic_1992}. This stochastic modification of the chain rule also lies at the heart of the derivation of the stochastic MB equations in Sec.~\ref{sec-variables}. For further recent examples of its usefulness in the investigation of light-matter interaction we refer to \cite{chuchurka_quantum_2023,chuchurka_stochastic_2023,chuchurka_stochastic2_2023}. Incorporating \eqref{eqn-ito-formula} into \eqref{eqn-sde-jc} yields an SDE of the form
\begin{equation}
  \diff\begin{bmatrix}
    \Phi_{t} \\
    \Sigma_{t}
  \end{bmatrix}=\begin{bmatrix}
    \function{\mat{A}_{\mathrm{JC}}}{\Phi_{t}} \\
    \function{\mat{a}}{\Phi_{t}}
  \end{bmatrix}\diff t+\begin{bmatrix}
    \function{\mat{B}_{\mathrm{JC}}}{\Phi_{t}} \\
    \function{\mat{b}}{\Phi_{t}}
  \end{bmatrix}\diff\mat{W}_{t}
  \label{eqn-sde-jc-tilde}
\end{equation}
for the expanded phase space variables $(\phi,\sigma)$ (note that the time-evolution of the $\phi$ part is unaffected because $\sigma$ does not couple back into $\phi$). The desired realization paths $\sigma_{n}^{r}$ can be computed directly from this SDE .
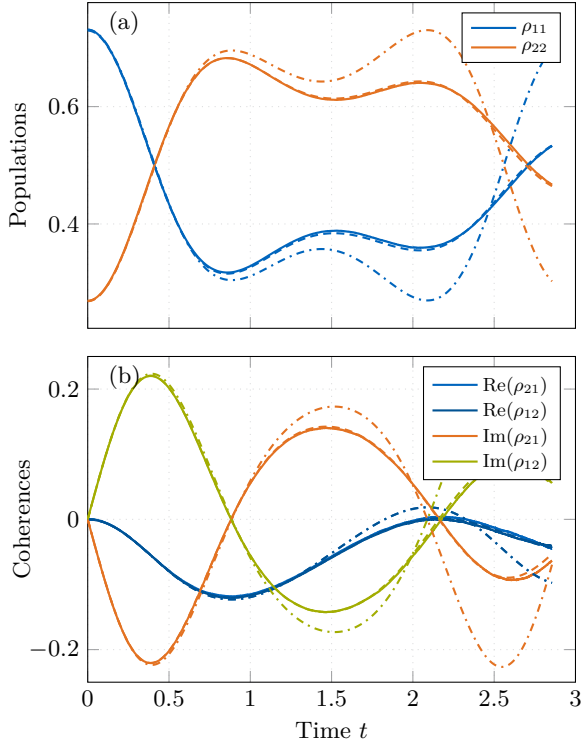
\begin{figure}[t]
  \centering
  \tikzexternalenable
  \begin{tikzpicture}
    \begin{groupplot}[
        group style={
            group size=1 by 2,
            vertical sep=0.375cm,
            xlabels at=edge bottom,
            xticklabels at=edge bottom,
            ylabels at=edge left,
          },
        width=0.45\textwidth,
        height=0.33\textwidth,
        xlabel={Time $t$},
        unit markings = parenthesis,
        x unit={\unit{\arbitraryunits}},
      ]
      \nextgroupplot[
        legend pos = north east,
        legend cell align = left,
        legend style={nodes={scale=0.8, transform shape}},
        xmin = 0,
        xmax = 3,
        ylabel = {Populations},
        y label style={at={(axis description cs:-0.1,.5)}, anchor=south},
        title = {(a)},
        title style={at={(0.025,0.95)}, anchor=north west},
        grid = both,
        grid style = {draw=TUMGrayLight, dotted, ultra thin},
      ]
      \addplot[color=TUMBlue, thick, solid] table [y=real_rho_11, x expr=\thisrow{t}*1000, col sep=comma] {fig-rho-sde.csv};
      \addplot[color=TUMOrange, thick, solid] table [y=real_rho_22, x expr=\thisrow{t}*1000, col sep=comma] {fig-rho-sde.csv};

      \addplot[color=TUMBlue, thick, dashdotted] table [y=real_rho_11, x expr=\thisrow{t}*1000, col sep=comma] {fig-mb.csv};
      \addplot[color=TUMOrange, thick, dashdotted] table [y=real_rho_22, x expr=\thisrow{t}*1000, col sep=comma] {fig-mb.csv};

      \addplot[color=TUMBlue, thick, densely dashed, forget plot] table [y=real_rho_11, x expr=\thisrow{t}*1000, col sep=comma] {fig-rho-n.csv};
      \addplot[color=TUMOrange, thick, densely dashed, forget plot] table [y=real_rho_22, x expr=\thisrow{t}*1000, col sep=comma] {fig-rho-n.csv};
      \legend{$\rho_{11}$, $\rho_{22}$}

      \nextgroupplot[
        legend pos = north east,
        legend cell align = left,
        legend style={nodes={scale=0.8, transform shape}},
        xmin = 0,
        xmax = 3,
        ymin = -0.25,
        ymax = 0.25,
        title = {(b)},
        title style={at={(0.025,0.95)}, anchor=north west},
        ylabel = {Coherences},
        y label style={at={(axis description cs:-0.1,.5)}, anchor=south},
        grid = both,
        grid style = {draw=TUMGrayLight, dotted, ultra thin},
      ]
      \addplot[color=TUMBlue, thick, solid] table [y=real_rho_21, x expr=\thisrow{t}*1000, col sep=comma] {fig-rho-sde.csv};
      \addplot[color=TUMBlueDark, thick, solid] table [y=real_rho_12, x expr=\thisrow{t}*1000, col sep=comma] {fig-rho-sde.csv};
      \addplot[color=TUMOrange, thick, solid] table [y=imag_rho_21, x expr=\thisrow{t}*1000, col sep=comma] {fig-rho-sde.csv};
      \addplot[color=TUMGreen, thick, solid] table [y=imag_rho_12, x expr=\thisrow{t}*1000, col sep=comma] {fig-rho-sde.csv};

      \addplot[color=TUMBlue, thick, dashdotted] table [y=real_rho_21, x expr=\thisrow{t}*1000, col sep=comma] {fig-mb.csv};
      \addplot[color=TUMBlueDark, thick, dashdotted] table [y=real_rho_12, x expr=\thisrow{t}*1000, col sep=comma] {fig-mb.csv};
      \addplot[color=TUMOrange, thick, dashdotted] table [y=imag_rho_21, x expr=\thisrow{t}*1000, col sep=comma] {fig-mb.csv};
      \addplot[color=TUMGreen, thick, dashdotted] table [y=imag_rho_12, x expr=\thisrow{t}*1000, col sep=comma] {fig-mb.csv};

      \addplot[color=TUMBlue, thick, densely dashed, forget plot] table [y=real_rho_21, x expr=\thisrow{t}*1000, col sep=comma] {fig-rho-n.csv};
      \addplot[color=TUMBlueDark, thick, densely dashed, forget plot] table [y=real_rho_12, x expr=\thisrow{t}*1000, col sep=comma] {fig-rho-n.csv};
      \addplot[color=TUMOrange, thick, densely dashed, forget plot] table [y=imag_rho_21, x expr=\thisrow{t}*1000, col sep=comma] {fig-rho-n.csv};
      \addplot[color=TUMGreen, thick, densely dashed, forget plot] table [y=imag_rho_12, x expr=\thisrow{t}*1000, col sep=comma] {fig-rho-n.csv};
      \legend{$\mathop{\mathrm{Re}}\left(\rho_{21}\right)$, $\mathop{\mathrm{Re}}\left(\rho_{12}\right)$,$\mathop{\mathrm{Im}}\left(\rho_{21}\right)$, $\mathop{\mathrm{Im}}\left(\rho_{12}\right)$}
    \end{groupplot}
  \end{tikzpicture}
  \tikzexternaldisable
  \caption{Time evolution of the atomic density operator. The diagonal entries $\rho_{11}$ (blue), $\rho_{22}$ (orange), and off-diagonal entries $\rho_{21}$ (real part blue, imaginary part orange), and $\rho_{12}$ (real part dark blue, imaginary part green) are shown in (a) and (b), respectively. The solid lines belong to the full-wave Jaynes-Cummings SDE simulation, the dashed lines to an independent truncated number state basis Heisenberg picture reference simulation, and the dash-dotted lines to a complementary semiclassical MB simulation without dissipation.\vspace{-2mm}}
  \label{fig-results}
\end{figure}

In this way, we can determine the time evolution of the coherences
\begin{equation}
  \begin{alignedat}{2}
     & \function{\rho_{21}}{t} &  & =\function{\braket{\op{S}_{-}}}{t}\, , \\
     & \function{\rho_{12}}{t} &  & =\function{\braket{\op{S}_{+}}}{t}\, ,
  \end{alignedat}
\end{equation}
and the population inversion
\begin{equation}
  \function{\nu}{t}=\function{\rho_{22}}{t}-\function{\rho_{11}}{t}=\function{\braket{2\op{S}_{z}}}{t}\, ,
\end{equation}
from \num{2995} SDE runs. Actually \num{3000} SDE runs were started, of which five diverged. Discarding divergent runs interferes with the statistics, so it can only be done on a small scale. Therefore, it is not a fix for significant stability issues.

The results are shown in \figref{fig-results}. It can be seen that our SDE solutions (solid) are in good agreement with the Heisenberg picture reference simulation (dashed). For comparison, we also depict a complementary semiclassical MB simulation (dash-dotted) utilizing \eqref{eqn-bloch-equations} without dissipation, i.e., $\gamma_{1}=\gamma_{2}=0$ and \eqref{eqn-maxwell-equations}, which only initially follows the full quantum solutions and quickly deviates. These vanishing relaxation rates are justified by the fact that the full-wave Jaynes-Cummings model does not include collisional dephasing, i.e., spontaneous emission is the only remaining cause for dissipation/fluctuation in the perfect cavity from the viewpoint of the Lindblad formalism with classical optical fields. However, it turns out that in our case, where the dipole moment is parallel to the mirrors and the half-wavelength of the atomic transition does not fit into the cavity, spontaneous emission is strongly suppressed~\cite{alves_spontaneous_2000}. Indeed, the phenomenological cavity-modified spontaneous emission rate derived by means of the Weisskopf-Wigner approximation equals zero~\cite{milonni_spontaneous_1973}. This unfavorable case for the traditional Heisenberg-Langevin approach demonstrates the benefits of using tools from quantum optics that avoid a priori semiclassical approximations in the derivation of the MB equations (see Sec.~\ref{sec-variables}) while yielding stochastic corrections that do not rely on the fluctuation-dissipation theorem. We note that the full-wave Jaynes-Cummings model from our SDE and semiclassical MB simulations in \figref{fig-results} can be applied to the physical situation of a single trapped cold atom in a detuned high-quality optical cavity with low-intensity optical fields~\cite{reiserer_cavity-based_2015}.

For time scales that are too long, positive $P$ representation SDEs for nonlinear quantum optical systems are known to sometimes produce wrong simulation results, and the causes and accompanying warning signs are analyzed in~\cite{gilchrist_positive_1997}. For example, the critical assumption in the derivation of the FPE, namely, that boundary terms from partial integration vanish (see Sec.~\ref{sec-sde}) can lose its validity in the course of the time evolution~\cite{carmichael_statistical_2_2007}. For some bosonic systems, drift gauges have been successfully used to remove these boundary terms and thereby extend the time interval that can be simulated correctly~\cite{deuar_gauge_2002,drummond_quantum_2003}. Our full-wave Jaynes-Cummings SDE simulation does not reach this natural time limit (if it exists) because stability issues set in beforehand. This is not particularly surprising given the singularities of the $\tanh$ terms in \eqref{eqn-drift-vector-jc-special}. Unfortunately, we were unable to find nonorthogonal fermionic basis states such that both the resulting drift vector and noise matrix are singularity-free. A more sophisticated SDE simulation might be able to circumvent this issue by employing a set of complementary nonorthogonal fermionic basis states and switching between them on the fly.
\vspace{-6mm}

\section{\label{sec-dissipation}Incorporation of Dissipation}

\vspace{-3mm}
It turns out that the dissipation model from Sec.~\ref{sec-semiclassical} is compatible with the nonorthogonal basis state expansion method for deriving SDEs, which was developed in Sec.~\ref{sec-sde}. This observation has the potential to pave the way for future, more realistic SDE simulations of light-matter interaction.

In order to check the validity of this claim, we calculate the effect of the three terms of the dissipation superoperator \eqref{eqn-dissipation-superoperator} on the projector $\op{\Lambda}=\op{\Lambda}_{\mathrm{JC}}$ from \eqref{eqn-lambda-jc}. Using the projector differential identities \eqref{eqn-fermionic-formulas} we obtain
\vspace{-2mm}
\begin{widetext}
  \vspace{-5mm}
  \begin{align}
    2\op{S}_{z}\op{\Lambda}\op{S}_{z}-\frac{1}{2}\op{\Lambda}                                                           & =\biggl[-\frac{h\bigl(1-h\tilde{h}\bigr)}{h'\bigl(1+h\tilde{h}\bigr)}\pdd{}{z}-\frac{\tilde{h}\bigl(1-h\tilde{h}\bigr)}{\tilde{h}'\bigl(1+h\tilde{h}\bigr)}\pdd{}{w}+\frac{1}{2}\biggl(\frac{2h\tilde{h}}{h'\tilde{h}'}\frac{\partial^2}{\partial z\partial w}+\frac{2h\tilde{h}}{h'\tilde{h}'}\frac{\partial^2}{\partial w\partial z}\biggr)\biggr]\op{\Lambda}\, ,\nonumber                    \\
    \op{S}_{-}\op{\Lambda}\op{S}_{+}-\frac{1}{2}\left(\op{S}_{z}\op{\Lambda}+\op{\Lambda}\op{S}_{z}+\op{\Lambda}\right) & =\biggl[-\frac{h\bigl(1+3h\tilde{h}\bigr)}{2h'\bigl(1+h\tilde{h}\bigr)}\pdd{}{z}-\frac{\tilde{h}\bigl(1+3h\tilde{h}\bigr)}{2\tilde{h}'\bigl(1+h\tilde{h}\bigr)}\pdd{}{w}+\frac{1}{2}\biggl(\frac{h^{2}\tilde{h}^{2}}{h'\tilde{h}'}\frac{\partial^2}{\partial z\partial w}+\frac{h^{2}\tilde{h}^{2}}{h'\tilde{h}'}\frac{\partial^2}{\partial w\partial z}\biggr)\biggr]\op{\Lambda}\, ,\nonumber  \\
    \op{S}_{+}\op{\Lambda}\op{S}_{-}+\frac{1}{2}\left(\op{S}_{z}\op{\Lambda}+\op{\Lambda}\op{S}_{z}-\op{\Lambda}\right) & =\biggl[\frac{h\bigl(3+h\tilde{h}\bigr)}{2h'\bigl(1+h\tilde{h}\bigr)}\pdd{}{z}+\frac{\tilde{h}\bigl(3+h\tilde{h}\bigr)}{2\tilde{h}'\bigl(1+h\tilde{h}\bigr)}\pdd{}{w}+\frac{1}{2}\biggl(\frac{1}{h'\tilde{h}'}\frac{\partial^2}{\partial z\partial w}+\frac{1}{h'\tilde{h}'}\frac{\partial^2}{\partial w\partial z}\biggr)\biggr]\op{\Lambda}\, .\label{eqn-differential-identities-dissipation}
    \vspace{-5mm}
  \end{align}
  As in \eqref{eqn-commutator-hamiltonian-lambda}, only partial derivatives with respect to the phase space variables of orders one and two occur. Therefore, these terms can be incorporated into the FPE described by the drift vector \eqref{eqn-drift-vector-jc} and diffusion matrix \eqref{eqn-diffusion-matrix-jc}. Note that owing to the form of the left-hand sides of these expressions, one would actually expect an atypical FPE featuring a potential term $V_{\mathrm{JC}+}$,
  \begin{equation}
    \begin{split}
      \pdd{P}{t}(\phi,t)= & \Biggl[\function{V_{\mathrm{JC}+}}{\phi}-\sum_{i=1}^{n}\pdd{}{\phi_{i}}\function{A_{\mathrm{JC}+,i}}{\phi}+\frac{1}{2}\sum_{i,j=1}^{n}\frac{\partial^2}{\partial\phi_{i}\partial\phi_{j}}\function{D_{\mathrm{JC}+,ij}}{\phi}\Biggr]\function{P}{\phi,t}\, ,
    \end{split}
    \label{eqn-fpe-potential}
  \end{equation}
  which cannot be translated into an SDE directly. Here, one needs to first introduce an additional phase space variable designed to remove the potential term~\cite{deuar_gauge_2002}. However, this has the drawback of not only increasing the computational load but also making the physical interpretation of the phase space variables more complicated. In our case, we are fortunate that because of a series of cancellations, $V_{\mathrm{JC}+}=0$ holds. From \eqref{eqn-differential-identities-dissipation}, it follows that the updated drift vector becomes
  \begin{equation}
    \function{\mat{A}_{\mathrm{JC}+}}{\phi}=\begin{bmatrix}
      -\i\omega_{1}\alpha_{1}-\i\function{g}{\omega_{1}}\sin(k_{1}x_{0})\frac{h+\tilde{h}}{1+h\tilde{h}}                                                                                                                                                                                                                                                          \\
      \i\omega_{1}\beta_{1}+\i\function{g}{\omega_{1}}\sin(k_{1}x_{0})\frac{h+\tilde{h}}{1+h\tilde{h}}                                                                                                                                                                                                                                                            \\
      \vdots                                                                                                                                                                                                                                                                                                                                                      \\
      -\i\omega_{N}\alpha_{N}-\i\function{g}{\omega_{N}}\sin(k_{N}x_{0})\frac{h+\tilde{h}}{1+h\tilde{h}}                                                                                                                                                                                                                                                          \\
      \i\omega_{N}\beta_{N}+\i\function{g}{\omega_{N}}\sin(k_{N}x_{0})\frac{h+\tilde{h}}{1+h\tilde{h}}                                                                                                                                                                                                                                                            \\
      -\i\Omega\frac{h}{h'}+\sum_{n=1}^{N}\i\function{g}{\omega_{n}}\left(\alpha_{n}+\beta_{n}\right)\sin(k_{n}x_{0})\frac{h^{2}-1}{h'}-r_{\mathrm{p}}\frac{h\left(1-h\tilde{h}\right)}{h'\left(1+h\tilde{h}\right)}-r_{21}\frac{h\left(1+3h\tilde{h}\right)}{2h'\left(1+h\tilde{h}\right)}+r_{12}\frac{h\left(3+h\tilde{h}\right)}{2h'\left(1+h\tilde{h}\right)} \\
      \i\Omega\frac{\tilde{h}}{\tilde{h}'}-\sum_{n=1}^{N}\i\function{g}{\omega_{n}}\left(\alpha_{n}+\beta_{n}\right)\sin(k_{n}x_{0})\frac{\tilde{h}^{2}-1}{\tilde{h}'}-r_{\mathrm{p}}\frac{\tilde{h}\left(1-h\tilde{h}\right)}{\tilde{h}'\left(1+h\tilde{h}\right)}-r_{21}\frac{\tilde{h}\left(1+3h\tilde{h}\right)}{2\tilde{h}'\left(1+h\tilde{h}\right)}+r_{12}\frac{\tilde{h}\left(3+h\tilde{h}\right)}{2\tilde{h}'\left(1+h\tilde{h}\right)}
    \end{bmatrix}\, .
  \end{equation}
\end{widetext}
Moreover, the updated diffusion matrix
\begin{equation}
  \function{\mat{D}_{\mathrm{JC}+}}{\phi}=\begin{bmatrix}
    \mat{0}                        & \cdots & \mat{0}                        & \i\function{\mat{D}_{1}}{\phi} \\
    \vdots                         & \ddots & \vdots                         & \vdots                         \\
    \mat{0}                        & \cdots & \mat{0}                        & \i\function{\mat{D}_{N}}{\phi} \\
    \i\function{\mat{D}_{1}}{\phi} & \cdots & \i\function{\mat{D}_{N}}{\phi} & \mat{D}(\phi)
  \end{bmatrix}
\end{equation}
contains the additional nonzero block
\begin{equation}
  \function{\mat{D}}{\phi}=d(\phi)\begin{bmatrix}
    0 & 1 \\
    1 & 0
  \end{bmatrix}\, ,
\end{equation}
with the entry
\begin{equation}
  \function{d}{\phi}=r_{\mathrm{p}}\frac{2\function{h}{z}\function{\tilde{h}}{w}}{\function{h'}{z}\function{\tilde{h}'}{w}}+r_{21}\frac{\function{h}{z}^{2}\function{\tilde{h}}{w}^{2}}{\function{h'}{z}\function{\tilde{h}'}{w}}+r_{12}\frac{1}{\function{h'}{z}\function{\tilde{h}'}{w}}\, .
\end{equation}
Thus, the additional quantum noise due to dissipation depends on the scattering and pure dephasing rates as well as the fermionic phase space variables. For the updated SDE, we need to again factorize $\mat{D}_{\mathrm{JC}+}$. To this end, consider
\begin{equation}
  \function{\mat{T}}{\phi}=\sqrt{\frac{\function{d}{\phi}}{2}}\begin{bmatrix}
    -\i & 1 \\
    \i  & 1
  \end{bmatrix}\, ,
\end{equation}
and set
\vspace{-1mm}
\begin{equation}
  \function{\mat{B}}{\phi}=\begin{bmatrix}
    \mat{0} & \cdots & \mat{0} & \mat{0}                  \\
    \vdots  & \ddots & \vdots  & \vdots                   \\
    \mat{0} & \cdots & \mat{0} & \mat{0}                  \\
    \mat{0} & \cdots & \mat{0} & \function{\mat{T}}{\phi}
  \end{bmatrix}\, ,
  \label{eqn-b}
  \vspace{-1mm}
\end{equation}
where the nonzero block is located in row $N+1$ and column $N+1$ of the block matrix. Then modifying \eqref{eqn-noise-matrix-jc} to
\begin{equation}
  \function{\mat{B}_{\mathrm{JC}+}}{\phi}=\begin{bmatrix}
    \sqrt{\i}\function{\mat{B}_{1}}{\phi} & \cdots & \sqrt{\i}\function{\mat{B}_{N}}{\phi} & \function{\mat{B}}{\phi}
  \end{bmatrix}\, ,
  \label{eqn-noise-matrix-jcp}
\end{equation}
suffices to satisfy the requirement $\mat{B}_{\mathrm{JC}+}\mat{B}_{\mathrm{JC}+}^{\transp}=\mat{D}_{\mathrm{JC}+}$.

\section{\label{sec-variables}The stochastic MB equations}

We are almost in the position to accomplish our main goal of using the SDE for the full-wave Jaynes-Cummings type system, including dissipation derived in Sec.~\ref{sec-dissipation} to gain new insight into the MB equations from Sec.~\ref{sec-semiclassical}. All that remains to be done in preparation is to bring this SDE into a more convenient form by two subsequent changes of variables.

First, consider the change of variables
\begin{equation}
  \phi\rightarrow\psi=(\alpha_{1},\beta_{1},\ldots,\alpha_{N},\beta_{N},\rho_{21},\rho_{12},\nu)\, ,
\end{equation}
with the new fermionic phase space variables
\begin{equation}
  \begin{alignedat}{2}
     & \rho_{21} &  & =\frac{\function{h}{z}}{1+\function{h}{z}\function{\tilde{h}}{w}}\, ,                          \\
     & \rho_{12} &  & =\frac{\function{\tilde{h}}{w}}{1+\function{h}{z}\function{\tilde{h}}{w}}\, ,                  \\
     & \nu       &  & =\frac{\function{h}{z}\function{\tilde{h}}{w}-1}{1+\function{h}{z}\function{\tilde{h}}{w}}\, .
  \end{alignedat}
\end{equation}
Their symbols are motivated by \eqref{eqn-expression-s-minus} and analogous results arising from $\op{S}_{+}$ and $\op{S}_{z}$. For later use we introduce the stochastic processes $\mathrm{P}_{21}$, $\mathrm{P}_{12}$, and $\mathrm{N}$ belonging to the new fermionic phase space variables. They are denoted by uppercase Greek letters (i.e., $\rho\leadsto\mathrm{P}$ and $\nu\leadsto\mathrm{N}$) in accordance with our naming convention (see Sec.~\ref{sec-sde}). Note that we have introduced an additional fermionic phase space dimension (see Sec.~\ref{sec-simulation}). This is a convenient choice for our purposes, but it is actually not necessary. It should also be kept in mind that there is no reason for realizations $\rho_{21}$ and $\rho_{12}$ being complex conjugates. The change back to the previous phase space variables $\psi\rightarrow\phi$ is determined by
\begin{equation}
  \begin{alignedat}{2}
     & \function{h}{z}         &  & =\frac{2\rho_{21}}{1-\nu}=\frac{1+\nu}{2\rho_{12}}\, , \\
     & \function{\tilde{h}}{w} &  & =\frac{1+\nu}{2\rho_{21}}=\frac{2\rho_{12}}{1-\nu}\, ,
  \end{alignedat}
\end{equation}
provided that $h$ and $\tilde{h}$ cooperate.

Second, consider the change of variables
\begin{equation}
  \psi\rightarrow\bar{\psi}=(\epsilon_{1},\eta_{1},\ldots,\epsilon_{N},\eta_{N},\rho_{21},\rho_{12},\nu)
\end{equation}
with the new bosonic phase space variables
\begin{equation}
  \begin{alignedat}{2}
     & \epsilon_{n} &  & =\beta_{n}+\alpha_{n},                \\
     & \eta_{n}     &  & =\i\left(\beta_{n}-\alpha_{n}\right).
  \end{alignedat}
  \label{eqn-epsilonn-mun}
\end{equation}
Clearly this is motivated by \eqref{eqn-en-hn} and the fact that due to the equalities $\braket{\langle\beta_{n}\rangle}=\braket{\op{a}_{n}^{\dagger}}$ and $\braket{\langle\alpha_{n}\rangle}=\braket{\op{a}_{n}}$ (see Sec.~\ref{sec-simulation}) it is consistent to replace $\braket{\op{a}_{n}^{\dagger}}$ and $\braket{\op{a}_{n}}$ with $\beta_{n}$ and $\alpha_{n}$, respectively. Again, the stochastic processes belonging to the new bosonic phase space variables are denoted by uppercase Greek letters $\mathrm{E}_{n}$ and $\mathrm{H}_{n}$ (i.e., $\epsilon\leadsto\mathrm{E}$ and $\eta\leadsto\mathrm{H}$). The expressions
\begin{equation}
  \begin{alignedat}{2}
     & \alpha_{n} &  & =\frac{\epsilon_{n}+\i\eta_{n}}{2}\, , \\
     & \beta_{n}  &  & =\frac{\epsilon_{n}-\i\eta_{n}}{2}
  \end{alignedat}
\end{equation}
occur in the change back to the previous phase space variables $\bar{\psi}\rightarrow\psi$.

An application of It\^{o}'s formula \eqref{eqn-ito-formula} shows that after these two changes of variables, the drift vector becomes
\begin{equation}
  \function{\bar{\mat{A}}_{\mathrm{JC}+}}{\bar{\psi}}=\scalebox{0.88}{$\displaystyle\begin{bmatrix}
        \omega_{1}\eta_{1}                                                                                             \\
        -\omega_{1}\epsilon_{1}-2\function{g}{\omega_{1}}\sin(k_{1}x_{0})\left(\rho_{21}+\rho_{12}\right)              \\
        \vdots                                                                                                         \\
        \omega_{N}\eta_{N}                                                                                             \\
        -\omega_{N}\epsilon_{N}-2\function{g}{\omega_{N}}\sin(k_{N}x_{0})\left(\rho_{21}+\rho_{12}\right)              \\
        -\i\Omega\rho_{21}+\sum_{n=1}^{N}\i\function{g}{\omega_{n}}\epsilon_{n}\sin(k_{n}x_{0})\nu-\gamma_{2}\rho_{21} \\
        \i\Omega\rho_{12}-\sum_{n=1}^{N}\i\function{g}{\omega_{n}}\epsilon_{n}\sin(k_{n}x_{0})\nu-\gamma_{2}\rho_{12}  \\
        2\sum_{n=1}^{N}\i\function{g}{\omega_{n}}\epsilon_{n}\sin(k_{n}x_{0})\left(\rho_{21}-\rho_{12}\right)-\gamma_{1}\left(\nu-\nu_{0}\right)
      \end{bmatrix}$}\, .
  \label{eqn-drift-vector-jcp-bar}
\end{equation}
It is interesting that the unexpected (from the viewpoint of the ordinary chain rule) noise-matrix-dependent contribution to the drift vector in It\^o's formula is responsible for the familiar form of the relaxation terms in the last three rows of this matrix [see \eqref{eqn-bloch-equations}]. The considerably more complicated case of the remaining noise matrix is treated in Appendix~\ref{sec-noise-matrix}. It turns out that all entries of $\function{\bar{\mat{B}}_{\mathrm{JC}+}}{\bar{\psi}}$ are problematic regarding numerical stability. The circumvention of such issues that stand in the way of successful simulations based on this approach requires further work. We limit ourselves to preparatory theoretical considerations as a necessary first step towards this goal.

Let us return to the investigation of the relation between the SDE
\begin{equation}
  \dd{\bar{\Psi}_{t}}{t}=\function{\bar{\mat{A}}_{\mathrm{JC}+}}{\bar{\Psi}_{t}}+\function{\bar{\mat{B}}_{\mathrm{JC}+}}{\bar{\Psi}_{t}}\bm\xi_{t}
  \label{eqn-sde-jcp-bar}
\end{equation}
and the MB equations. We want to derive a stochastic semiclassical scheme that stays as close as possible to this SDE. For this purpose, we need to replace the bosonic phase space variables with stochastic electric and magnetic fields. Their time evolution equations
\vspace{-1mm}
\begin{equation}
  \begin{split}
    \dd{\mathrm{E}_{n,t}}{t}= & \,\omega_{n}\mathrm{H}_{n,t}+\left[\function{\bar{\mat{B}}_{\mathrm{JC}+}}{\bar{\Psi}_{t}}\bm\xi_{t}\right]_{2n-1}\, , \\
    \dd{\mathrm{H}_{n,t}}{t}= & -\omega_{n}\mathrm{E}_{n,t}-2g(\omega_n)\left(\mathrm{P}_{21,t}+\mathrm{P}_{12,t}\right)\sin(k_{n}x_{0})               \\
                              & +\left[\function{\bar{\mat{B}}_{\mathrm{JC}+}}{\bar{\Psi}_{t}}\bm\xi_{t}\right]_{2n}
  \end{split}
  \label{eqn-maxwell-equations-modes-sde}
  \vspace{-2mm}
\end{equation}
[see \eqref{eqn-drift-vector-jcp-bar}] correspond to the cavity mode form of the 1D~Maxwell equations \eqref{eqn-maxwell-equations-modes-qm}. Here the notation $\left[\vec{v}\right]_{n}$ picks the $n$th entry of a vector $\vec{v}$. Consequently, a comparison with \eqref{eqn-fields-qm} shows that it is consistent to regard
\vspace{-1mm}
\begin{equation}
  \begin{alignedat}{2}
     & \function{\mathrm{E}}{x,t} &  & =\sum_{n=1}^{N}\function{e_{\mathrm{p}}}{\omega_{n}}\mathrm{E}_{n,t}\sin(k_{n}x)\, ,         \\
     & \function{\mathrm{H}}{x,t} &  & =-\frac{1}{Z}\sum_{n=1}^{N}\function{e_{\mathrm{p}}}{\omega_{n}}\mathrm{H}_{n,t}\cos(k_{n}x)
  \end{alignedat}
  \label{eqn-fields-sde}
  \vspace{-2mm}
\end{equation}
as the required stochastic electric and magnetic fields. With the help of \eqref{eqn-maxwell-equations-modes-sde} and \eqref{eqn-fields-sde}, we obtain
\vspace{-1mm}
\begin{align}
  \mu_{0}\pdd{\mathrm{H}}{t}(x,t)=         & \,\pdd{\mathrm{E}}{x}(x,t)+\frac{2}{c}\sum_{n=1}^{N}\function{e_{\mathrm{p}}}{\omega_{n}}\function{g}{\omega_n}\left(\mathrm{P}_{21,t}+\mathrm{P}_{12,t}\right)\nonumber                    \\
                                           & \times\sin(k_{n}x_{0})\cos(k_{n}x)-\frac{1}{c}\sum_{n=1}^{N}\function{e_{\mathrm{p}}}{\omega_{n}}\nonumber                                                                                  \\
                                           & \times\left[\function{\bar{\mat{B}}_{\mathrm{JC}+}}{\bar{\Psi}_{t}}\bm\xi_{t}\right]_{2n}\cos(k_{n}x)\, ,\nonumber                                                                          \\
  \varepsilon_{0}\pdd{\mathrm{E}}{t}(x,t)= & \,\pdd{\mathrm{H}}{x}(x,t)+\varepsilon_{0}\sum_{n=1}^{N}\function{e_{\mathrm{p}}}{\omega_{n}}\left[\function{\bar{\mat{B}}_{\mathrm{JC}+}}{\bar{\Psi}_{t}}\bm\xi_{t}\right]_{2n-1}\nonumber \\
                                           & \times\sin(k_{n}x)\, ,\label{eqn-maxwell-equations-stochastic}
  \vspace{-2mm}
\end{align}
a new version of the 1D~Maxwell equations with quantum noise corrections and an unexpected light-matter interaction term, which occurs in the first and not the second equation [see \eqref{eqn-maxwell-equations}]. This can be explained by rewriting the 1D~Maxwell equations in terms of the electric displacement field $D_{z}=\varepsilon_{0}E_{z}+P_{z}$, i.e.,
\vspace{-1mm}
\begin{equation}
  \begin{alignedat}{2}
     & \mu_{0}\pdd{H_{y}}{t} &  & =\frac{1}{\varepsilon_{0}}\left(\pdd{D_{z}}{x}-\pdd{P_{z}}{x}\right)\, , \\
     & \pdd{D_{z}}{t}        &  & =\pdd{H_{y}}{x}\, .
  \end{alignedat}
  \vspace{-2mm}
\end{equation}
Here the polarization also occurs in the first equation, and the derivative is spatial instead of temporal, which is compatible with the factor
\vspace{-1mm}
\begin{equation}
  \cos(k_{n}x)=\frac{1}{k_{n}}\dd{\sin(k_{n}x)}{x}
  \vspace{-2mm}
\end{equation}
in the light-matter interaction term of \eqref{eqn-maxwell-equations-stochastic}. This means that the electric field operator $\op{E}_{z}$ in \eqref{eqn-field-operators-qm} in fact corresponds to the operator $\op{D}_{z}/\varepsilon_{0}$ because it implicitly contains the light-matter interaction [the time evolution of $\function{\op{a}_{n}^{\dagger}}{t}$ and $\function{\op{a}_{n}}{t}$ depends on $\function{\op{S}_{z}}{t}$, $\function{\op{S}_{+}}{t}$, and $\function{\op{S}_{-}}{t}$]. Some related observations concerning the connection between second quantization and Maxwell's equations in nonlinear optics albeit not in the present SDE context can be found in~\cite{quesada_electric_2017}.

In order to be able to eliminate the bosonic phase space variables from the rest of the SDE, we need to postulate that the bosonic-fermionic coupling constants are related to the dipole moment from \eqref{eqn-dipole-operator} via the formula
\begin{equation}
  m_{21}=-\frac{\hbar g(\omega_{n})}{e_{\mathrm{p}}(\omega_{n})}\, .
  \label{eqn-dipole-coupling}
\end{equation}
This relation can also be found in~\cite{rand_lectures_2016}, where it is derived by other means. Then \eqref{eqn-fields-sde} together with \eqref{eqn-dipole-coupling} yields
\begin{equation}
  \sum_{n=1}^{N}\i\function{g}{\omega_{n}}\mathrm{E}_{n,t}\sin(k_{n}x_{0})=-\frac{\i}{\hbar}m_{21}\function{\mathrm{E}}{x_{0},t}\, .\label{eqn-dipole-coupling-consequence}
\end{equation}
Using \eqref{eqn-drift-vector-jcp-bar} and \eqref{eqn-dipole-coupling-consequence}, the time evolution equations for the fermionic phase space variables become
\begin{align}
  \dd{\mathrm{P}_{21,t}}{t}= & \,-\i\Omega\mathrm{P}_{21,t}-\frac{\i}{\hbar}m_{21}\function{\mathrm{E}}{x_{0},t}\mathrm{N}_{t}-\gamma_{2}\mathrm{P}_{21,t}\nonumber                            \\
                             & +\left[\function{\bar{\mat{B}}_{\mathrm{JC}+}}{\bar{\Psi}_{t}}\bm\xi_{t}\right]_{2N+1}\, ,\nonumber                                                             \\
  \dd{\mathrm{P}_{12,t}}{t}= & \,\i\Omega\mathrm{P}_{12,t}+\frac{\i}{\hbar}m_{21}\function{\mathrm{E}}{x_{0},t}\mathrm{N}_{t}-\gamma_{2}\mathrm{P}_{12,t}\nonumber                             \\
                             & +\left[\function{\bar{\mat{B}}_{\mathrm{JC}+}}{\bar{\Psi}_{t}}\bm\xi_{t}\right]_{2N+2}\, ,\nonumber                                                             \\
  \dd{\mathrm{N}_{t}}{t}     & \,=2\frac{\i}{\hbar}m_{21}\function{\mathrm{E}}{x_{0},t}\left(\mathrm{P}_{12,t}-\mathrm{P}_{21,t}\right)-\gamma_{1}\left(\mathrm{N}_{t}-\nu_{0}\right)\nonumber \\
                             & +\left[\function{\bar{\mat{B}}_{\mathrm{JC}+}}{\bar{\Psi}_{t}}\bm\xi_{t}\right]_{2N+3}\, ,
  \label{eqn-bloch-equations-stochastic}
\end{align}
which are just the Bloch equations \eqref{eqn-bloch-equations} with added quantum noise. Combining the results \eqref{eqn-maxwell-equations-stochastic} and \eqref{eqn-bloch-equations-stochastic} yields an SPDE of a form described in~\cite{werner_robust_1997}, corresponding to the MB equations. Note that the quantum noise corrections and the light-matter interaction term in the new version of the 1D~Maxwell equations bear the last remaining trace of the original cavity mode-dependent ansatz, which was the starting point of our derivation.

Clearly, before this SPDE can be used in practice, a consistent numerical treatment compatible with the FDTD method and leveraging its numerical efficiency needs to be specified. To accomplish this goal, it might be necessary to identify and keep only the most important quantum noise terms in order to obtain a stable stochastic semiclassical scheme for the future simulation of nonclassical light in optoelectronic devices.

\section{\label{sec-conclusion}Conclusion}

In this paper, we present a different way to improve the usefulness of the positive $P$ representation for fermionic or mixed bosonic and fermionic systems by highlighting the importance of choosing the nonorthogonal fermionic basis states adequately. Additionally, we fully describe the initialization of the corresponding fermionic phase space variables and provide a simple example simulation of the full-wave Jaynes-Cummings SDE as a starting point for future applications in quantum optics. With the help of these states, we identify the complex structure of the inherent quantum noise for a dissipative TLA in a perfect optical cavity, which enables the derivation of a stochastic correction to the MB equations in the form of an SPDE. In this way, we shed new light on how quantum optical phase space methods can be used to reveal and exploit the connection between semiclassical and field-quantized models for light-matter interaction. Despite the assumption of a perfect optical cavity, these results should not be limited to this case since, in the MB framework, mirror losses are easily treated by attenuating the optical field via an equivalent distributed conductivity. This is attractive because the actual modes of a lossy cavity are difficult to determine from second quantization~\cite{viviescas_field_2003}. Moreover, by disregarding the quantum noise terms in the SPDE, we obtain as a bonus a new treatment of the polarization in the MB equations that may be advantageous in the strong coupling and few photon regimes.

\begin{acknowledgments}
  The authors acknowledge the financial support by Deutsche Forschungsgemeinschaft (DFG) under the DFG DACH project (Project No.\ 471080402), by the FWF Project No.\ I~5682: \enquote{Cavity-assisted non-classical light generation,} and by the European Space Agency (ESA) Discovery EISI Project No.\ I-2023-02655: \enquote{Simulation toolbox for unconditionally secure on-chip satellite quantum communication networks operating in the telecom wavelength range.}
\end{acknowledgments}

\appendix

\section{\label{sec-initialization}\MakeUppercase{Initialization of the fermionic phase space variables}}

Consider a valid initial atomic density operator
\begin{equation}
  \op{\rho}_{\mathrm{A}}=\rho_{11}\ketbra{\downarrow}{\downarrow}+\rho_{12}\ketbra{\downarrow}{\uparrow}+\rho_{21}\ketbra{\uparrow}{\downarrow}+\rho_{22}\ketbra{\uparrow}{\uparrow}
\end{equation}
[see \eqref{eqn-matrix-rho}]. The construction of the associated probability distribution $P$ described below relies on all the properties of $\op{\rho}_{\mathrm{A}}$. Let us write $\rho_{11}=p$ and $\rho_{12}=r\e^{\i\phi}$ with $r\geq0$. By the hermiticity ${\op{\rho}_{\mathrm{A}}}^{\dagger}=\op{\rho}_{\mathrm{A}}$ and the trace condition $\trace\op{\rho}_{A}=1$ we have $\rho_{22}=1-p$ and $\rho_{21}=r\e^{-\i\phi}$. From the positive semidefiniteness $\op{\rho}_{\mathrm{A}}\geq0$ it follows that $0\leq p\leq1$ and $r^{2}\leq p(1-p)$. The stronger assumption $0<p<1$ is necessary for our purposes. Then
\begin{equation}
  q=\frac{r\left(1+K^{2}\right)}{K}\quad\text{with}\quad K=\sqrt{\frac{1}{p}-1}
\end{equation}
satisfies $0\leq q\leq1$. Hence $q_{1}=q$ and $q_{2}=q_{3}=(1-q)/2$ are probabilities that add up to $1$. Provided that there are $z_1,z_2,z_3\in\Cnums$ and $w_1,w_2,w_3\in\Cnums$ with
\begin{equation}
  \begin{alignedat}{6}
     & \function{h}{z_1}         &  & =K\e^{-\i\phi}\, ,\quad &  & \function{h}{z_2}         &  & =K\, ,\quad &  & \function{h}{z_3}         &  & =-K\, , \\
     & \function{\tilde{h}}{w_1} &  & =K\e^{\i\phi}\, ,\quad  &  & \function{\tilde{h}}{w_2} &  & =K\, ,\quad &  & \function{\tilde{h}}{w_3} &  & =-K\, ,
  \end{alignedat}
  \label{eqn-image-h}
\end{equation}
our construction together with \eqref{eqn-fermionic-lambda} yields
\begin{equation}
  \op{\rho}_{\mathrm{A}}=\sum_{i=1}^{3}q_{i}\function{\op{\Lambda}_{\mathrm{A}}}{z_{i},w_{i}}\, ,
\end{equation}
which means that the initial probability distribution
\begin{equation}
  \function{P}{z,w}=\sum_{i=1}^{3}q_{i}\dirac{z-z_{i}}\dirac{w-w_{i}}
\end{equation}
fits the requirements. In view of the mapping behavior of analytic functions (e.g., Picard's Theorem~\cite{krantz_complex_2004}), the theoretical chances for being able to satisfy \eqref{eqn-image-h} are excellent.

\section{\label{sec-noise-matrix}\MakeUppercase{Noise matrix for the phase space variables \texorpdfstring{$\bm{\bar{\psi}}$}{$\bar{\psi}$}}}

Each block of the noise matrix
\begin{equation}
  \function{\bar{\mat{B}}_{\mathrm{JC}+}}{\bar{\psi}}=\begin{bmatrix}
    \sqrt{\i}\function{\bar{\mat{B}}_{1}}{\bar{\psi}} & \cdots & \sqrt{\i}\function{\bar{\mat{B}}_{N}}{\bar{\psi}} & \function{\bar{\mat{B}}}{\bar{\psi}}
  \end{bmatrix}
  \label{eqn-noise-matrix-jcp-bar}
\end{equation}
[see \eqref{eqn-noise-matrix-jcp}] has the same structure as its counterpart before the change of variables [see \eqref{eqn-bn} and \eqref{eqn-b}]. The only difference is that $\bar{\mat{P}}_{n}(\bar{\psi})$, $\bar{\mat{Q}}_{n}(\bar{\psi})$, $\bar{\mat{R}}_{n}(\bar{\psi})$, $\bar{\mat{S}}_{n}(\bar{\psi})$, and $\bar{\mat{T}}(\psi)$ to be given below replace its blocks $\mat{P}_{n}(\phi)$, $\mat{Q}_{n}(\phi)$, $\mat{R}_{n}(\phi)$, $\mat{S}_{n}(\phi)$, and $\mat{T}(\phi)$, respectively. For the sake of simplicity, we restrict ourselves to the special case $\function{h}{z}=z$ (implying $\function{\tilde{h}}{w}=w$) which corresponds to the coherent spin states \eqref{eqn-coherent-spin-states}. Then it turns out that
\begin{equation}
  \begin{alignedat}{2}
     & \function{\bar{\mat{P}}_{n}}{\bar{\psi}} &  & =\sqrt{\frac{\function{g}{\omega_{n}}\sin(k_{n}x_{0})}{2}}\begin{bmatrix}
                                                                                                                 \i p & -p   \\
                                                                                                                 p    & \i p \\
                                                                                                               \end{bmatrix}\, , \\
     & \function{\bar{\mat{Q}}_{n}}{\bar{\psi}} &  & =\sqrt{\frac{\function{g}{\omega_{n}}\sin(k_{n}x_{0})}{2}}\begin{bmatrix}
                                                                                                                 -\i q & -q    \\
                                                                                                                 q     & -\i q \\
                                                                                                               \end{bmatrix}    \\
  \end{alignedat}
\end{equation}
holds, with the entries given by
\begin{equation}
  \begin{split}
    p & =\sqrt{\frac{4\rho_{21}^{2}}{\left(1-\nu\right)^{2}}-1}\, , \\
    q & =\sqrt{\frac{4\rho_{12}^{2}}{\left(1-\nu\right)^{2}}-1}\, .
  \end{split}
\end{equation}
The remaining blocks $\bar{\mat{R}}_{n}(\bar{\psi})$, $\bar{\mat{S}}_{n}(\bar{\psi})$, and $\bar{\mat{T}}(\bar{\psi})$ have three rows instead of two, as is the case for $\mat{R}_{n}(\phi)$, $\mat{S}_{n}(\phi)$, and $\mat{T}(\phi)$, because of the additional fermionic phase space dimension.

\noindent Indeed, we have
\vspace{-1mm}
\begin{equation}
  \begin{alignedat}{2}
     & \function{\bar{\mat{R}}_{n}}{\bar{\psi}} &  & =\sqrt{\frac{\function{g}{\omega_{n}}\sin(k_{n}x_{0})}{2}}\begin{bmatrix}
                                                                                                                 -\i r_{1} & -r_{1} \\
                                                                                                                 -\i r_{2} & -r_{2} \\
                                                                                                                 -\i r_{3} & -r_{3}
                                                                                                               \end{bmatrix}\, , \\
     & \function{\bar{\mat{S}}_{n}}{\bar{\psi}} &  & =\sqrt{\frac{\function{g}{\omega_{n}}\sin(k_{n}x_{0})}{2}}\begin{bmatrix}
                                                                                                                 -\i s_{1} & s_{1} \\
                                                                                                                 -\i s_{2} & s_{2} \\
                                                                                                                 -\i s_{3} & s_{3}
                                                                                                               \end{bmatrix}\, ,
  \end{alignedat}
\end{equation}
with the entries
\vspace{-1mm}
\begin{align*}
  r_{1} & =\frac{\left(1-\nu\right)^{2}}{4}\sqrt{\frac{4\rho_{21}^{2}}{\left(1-\nu\right)^{2}}-1}\, , \\
  r_{2} & =-\rho_{12}^{2}\sqrt{\frac{\left(1+\nu\right)^{2}}{4\rho_{12}^{2}}-1}\, ,                   \\
  r_{3} & =\rho_{12}\left(1-\nu\right)\sqrt{\frac{\left(1+\nu\right)^{2}}{4\rho_{12}^{2}}-1}\, ,      \\
  s_{1} & =-\rho_{21}^{2}\sqrt{\frac{\left(1+\nu\right)^{2}}{4\rho_{21}^{2}}-1}\, ,                   \\
  s_{2} & =\frac{\left(1-\nu\right)^{2}}{4}\sqrt{\frac{4\rho_{12}^{2}}{\left(1-\nu\right)^{2}}-1}\, , \\
  s_{3} & =\rho_{21}\left(1-\nu\right)\sqrt{\frac{\left(1+\nu\right)^{2}}{4\rho_{21}^{2}}-1}\, .
\end{align*}
Finally, the block arising from the incorporation of dissipation is
\vspace{-1mm}
\begin{equation}
  \function{\bar{\mat{T}}}{\bar{\psi}}=\sqrt{\frac{2r_{\mathrm{p}}\frac{1+\nu}{1-\nu}+r_{21}\left(\frac{1+\nu}{1-\nu}\right)^{2}+r_{12}}{2}}\begin{bmatrix}
    t_{11} & t_{12} \\
    t_{21} & t_{22} \\
    t_{31} & t_{32}
  \end{bmatrix}\, ,
\end{equation}
with the entries
\vspace{-1mm}
\begin{align*}
  t_{11} & =-\i\left[\rho_{21}^{2}+\frac{(1-\nu)^{2}}{4}\right]\, , \\
  t_{12} & =-\rho_{21}^{2}+\frac{(1-\nu)^{2}}{4}\, ,                \\
  t_{21} & =\i\left[\rho_{12}^{2}+\frac{(1-\nu)^{2}}{4}\right]\, ,  \\
  t_{22} & =-\rho_{12}^{2}+\frac{(1-\nu)^{2}}{4}\, ,                \\
  t_{31} & =\i\left(-\rho_{12}+\rho_{21}\right)(1-\nu)\, ,          \\
  t_{32} & =\left(\rho_{12}+\rho_{21}\right)(1-\nu)\, .             \\
\end{align*}

\bibliographystyle{apsrev4-2}
\bibliography{literature}

\end{document}